\documentclass{aa}  
\usepackage{natbib}
\bibpunct{(}{)}{;}{a}{}{,} 

\usepackage[T1]{fontenc}
\usepackage{graphicx}
\usepackage{subfigure}
\usepackage{epstopdf}
\usepackage{gensymb}
\usepackage{siunitx}
\usepackage[american]{circuitikz}
\usetikzlibrary{shapes,arrows}
\usepackage{mathtools}
\usepackage{commath}
\usepackage{diagbox}
\usepackage{makecell}
\usepackage{tabularx}

\begin{document}

   \title{Sensitivity of a Low-Frequency Polarimetric Radio Interferometer}


   \author{A. T. Sutinjo
          \inst{1}
          \and
           M. Sokolowski
          \inst{1}
           \and
           M. Kovaleva
          \inst{1}
          \and
          D. C. X. Ung
          \inst{1}
          \and
          J. W. Broderick
          \inst{1}
          \and
         R. B. Wayth
          \inst{1}
           \and
          D. B. Davidson
          \inst{1}
           \and
          S. J. Tingay
          \inst{1}
          }

   \institute{International Centre for Radio Astronomy Research (ICRAR), Curtin University, 6102 Australia\\
              \email{adrian.sutinjo@curtin.edu.au}
             }

   \date{
   Accepted by Astronomy and Astrophysics on 15 December 2020.
   }

 
  \abstract
  {The sensitivity of a radio interferometer is a key figure of merit (FoM) for a radio telescope. The sensitivity of a single polarized interferometer is typically given as antenna effective area over system temperature, $A_e/T_{sys}$, assuming an unpolarized source. For a dual-polarized polarimetric interferometer intended to observe sources of unknown polarization, the state of polarization must not be assumed \emph{a priori}. Furthermore, in contrast to the narrow field of view (FoV) of dish-based interferometers, the sensitivity of a polarimetric low-frequency radio interferometer warrants a careful review because of the very wide FoV of the dual-polarized antennas. A revision of this key FoM is particularly needed in the context of the Low-Frequency Square Kilometre Array (SKA-Low) where the sensitivity requirements are currently stated using $A_e/T_{sys}$ for single-polarized antenna system, which produces an ambiguity for off-zenith angles.} 
   {This paper aims to derive an expression for the sensitivity of a polarimetric radio interferometer that is valid for all-sky observations of arbitrarily polarized sources, with neither a restriction on FoV nor with any \emph{a priori} assumption regarding the polarization state of the source. We verify the resulting formula with an all-sky observation using the Murchison Widefield Array (MWA) telescope.}
   {The sensitivity expression is developed from first principles by applying the concept of System Equivalent Flux Density (SEFD) to a polarimetric radio interferometer (\emph{not} by computing $A_e/T_{sys}$). The SEFD is calculated from the standard deviation of the noisy flux density estimate for a target source due to system noise.}
   {The SEFD for a polarimetric radio interferometer is generally not  $1/\sqrt{2}$ of a single-polarized interferometer as is often assumed for narrow FoV. This assumption can lead to significant errors for a dual-polarized dipole based system, which is common in low-frequency radio astronomy: up to $\sim 15\%$ for a zenith angle (ZA) coverage of 45\degree, and up to $\sim45\%$ for 60\degree coverage.
   The worst case errors occur in the diagonal planes of the dipole for very wide FoV. This is demonstrated through theory, simulation and observations. Furthermore, using the resulting formulation, calculation of the off-zenith sensitivity is straightforward and unambiguous.}
   {For wide FoV observations pertinent to low-frequency radio interferometer such as the SKA-Low, the narrow FoV and the single-polarized sensitivity expressions are not correct and should be replaced by the formula derived in this paper.}

   \keywords{Instrumentation: interferometers--Techniques: polarimetric--Methods: analytical--
Methods: data analysis--Methods: numerical--
Methods: observational--Telescopes 
               }

   \maketitle
%
\section{Introduction}
\label{sec:intro}
A polarimetric radio interferometer consists of at least two dual-polarized antennas separated by a distance, as shown in~Fig.~\ref{fig:dual_pol}. The treatment of radio interferometry in the context of dish antenna-based systems is quite vast as summarized in~\citet{Thompson2017_intro}.  However, a low-frequency ($\lesssim \SI{350}{\mega\hertz}$) radio interferometer~\citep{2013PASA...30....7T, 2013A&A...556A...2V, 5109716, 8105424} notably differs from the dish antennas in that the low-frequency antennas are fixed with respect to the ground (i.e. they do not mechanically point to the target) and the telescope field of view (FoV) is wide. This difference is particularly evident in the treatment of polarimetric observation~\citep{doi:10.1002/2014RS005517, sokolowski_colegate_sutinjo_2017}.

Sensitivity is a metric for the faintest signal that can be measured by the telescope. In the context of radio astronomy, a conceptually meaningful measure is the \emph{system equivalent flux density} (SEFD). We may think of the SEFD as the flux of the target source that equals the standard deviation of the noisy flux density estimate  generated by the system. The conventional references for the sensitivity of a radio interferometer are~\cite{1989ASPC....6..139C, 1999ASPC..180..171W}. The derivations in these references are based on a single polarized dish antennas, in particular the Very Large Array (VLA) and Very-Long-Baseline Interferometry (VLBA). The result are extrapolated to the dual-polarized system via a statistical argument, i.e., the sensitivity is $1/\sqrt{2}$ times that of a single polarized system~\citep{1989ASPC....6..139C}. We shall demonstrate that this is correct for the low-frequency interferometer \emph{only} under certain conditions.

The need to derive the correct sensitivity metric for a radio polarimeter is being recognized~\citep{4414351, 8878981, Warnick_6018280}. Proper treatment of this is particularly important in next-generation low-frequency radio telescopes such as the low-frequency component of the Square Kilometre Array \citep{7928622} with very wide FoV and dual-polarized antennas with generally unequal responses at angles away from zenith. However, current requirements are still being stated using  antenna effective area over system temperature ($A_e/T_{sys}$) which refers to a single-polarized antenna system~\citep{SKA1BaselineDesign, SKA1L1req} assuming an unpolarized source. This leads to an ambiguity in the interpretation of the off-zenith sensitivity requirement. The work in~\cite{4414351} generalizes the concept of antenna temperature to Stokes parameters, which is a partial advance to $A_e/T_{sys}$ for Stokes parameters. However, as we shall discuss, the SEFD approach is more straightforward; the result can be extended with caution to a comparable metric to $A_e/T_{sys}$. Discussions of SEFD for Stokes parameters has been initiated in~\cite{8878981, Warnick_6018280}, but yet to be developed for the interferometer. Furthermore, in the context of the global collaboration towards SKA-Low, we seek an expression that has clear links to radio astronomy, and for which design implications may be readily inferred by engineers.

A polarimetric radio interferometer can estimate the flux density of a target source because of the dual-polarized antennas. Therefore, we expect to derive the SEFD directly for \emph{arbitrary polarization} without starting with $A_e/T_{sys}$, for which one has to assume something about the polarization of the incoming wave~\citep{4066948_Ko}. Strictly speaking, SEFD \emph{cannot} be defined for a \emph{single}-polarized radio telescope as it is \emph{unable} to measure the flux density of a target source due to insufficient information. The best an observer can do in this case is \emph{assume} that the target source is \emph{unpolarized} and apply a correction factor based on that assumption. This obviously does not apply to a target source of \emph{unknown} polarization and hence warrants a close review in Sec.~\ref{sec:back}.

\section{Background and aim}
\label{sec:back}

\begin{figure}[t]
\centering
\noindent
  \includegraphics[width=3.25in]{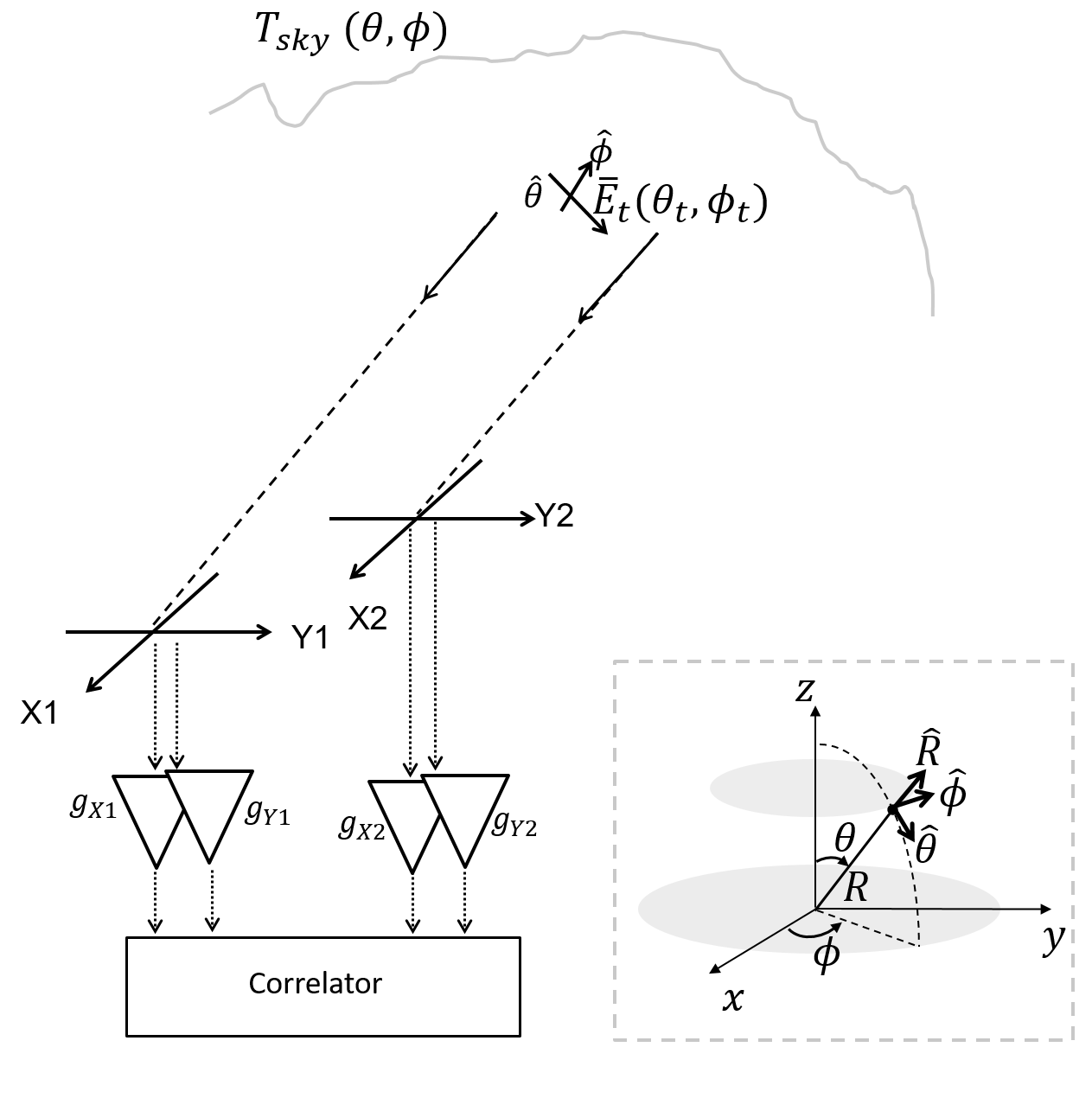}
\caption{Two dual-polarized low-frequency antennas forming an interferometer. The signal of interest is the partially polarized field $\bar{E}_t$ which is observed with a background noise $T_{sky}$. These signals combine to produce voltages at the antenna ports which are amplified and fed to the correlator. The inset shows the standard spherical coordinate system which we follow here.  Note that for a receiving antenna, the incident field propagates in $-\hat{R}$ direction such that the cross product relation is $\hat{\phi}\times \hat{\theta} = -\hat{R}$.}
\label{fig:dual_pol}
\end{figure}

The interferometer under consideration is depicted in Fig.\ref{fig:dual_pol}. Two dual-polarized antennas separated by a certain distance on a co-planar surface are illuminated by a background sky which we assume to be unpolarized with noise temperature distribution $T_{sky}(\theta,\phi)$ in the spherical coordinate system. A partially polarized source $\bar{E}_t(\theta_t,\phi_t)$ whose polarization property is unknown but is of interest is the target of observation. The study of polarization (polarimetry) is facilitated by the Stokes parameters~\citep{Wilson2009}, which are formed by linear combinations of the correlator output. 

The observation of $\bar{E}_t$ must contend with the undesired noise produced by $T_{sky}$ and the amplifiers in Fig.~\ref{fig:dual_pol}. We call the combination of these undesired noise sources, the system noise. The system noise is the dominant source of random error in the observation~\citep{1999ASPC..180..171W}. The higher the system noise relative to the antenna response in the direction of the target source, the higher the uncertainty in the resulting polarization estimates, i.e., the sensitivity of the instrument to $\bar{E}_t$ is low and vice versa.  

The aim of this work is to quantify the sensitivity of a radio interferometer for polarimetry of an arbitrarily located target signal of any polarization state. The SEFD~\citep{1999ASPC..180..171W} is the natural choice for the sensitivity metric because any source, be it fully polarized, partially polarized or unpolarized, has an unambiguous flux density ($\propto\norm{\bar{E}_t}^2=\left<|E_{t\theta}|^2 \right>+\left<|E_{t\phi}|^2 \right>$) that is \emph{independent} of its polarization state and the instrumentation used to observe it. Due to the dual-polarized antennas, the radio interferometer is suitable to detect the target source flux density. For comparison to the SEFD, consider $A_e/T_{sys}$, which is often used as the figure of merit (FoM) of a radio telescope.  The antenna effective area, $A_e$, is defined for matched polarization~\citep{6758443}. The practice of using $A_e/T_{sys}$ in radio astronomy implicitly \emph{assumes} an unpolarized source, and thereby the polarization mismatch factor of $1/2$~\citep{4066948_Ko} is attached to $A_e$~\citep{Thompson2017_ch1},
\begin{eqnarray}
\text{SEFD}|_{unpol.}=k\frac{T_{sys}}{\frac{1}{2}A_e}=2k\frac{T_{sys}}{A_e}.
\label{eqn:SEFD_AonT_formula}
\end{eqnarray}
Clearly, this does \emph{not} apply to polarimetry which concerns an unknown polarization state. 

The system equivalent noise contribution to sensitivity is conceptually simple. Consider the following thought experiment that is consistent with the approach outlined for the sensitivity calculation of a radio interferometer in~\cite{1999ASPC..180..171W}. Suppose we are able to turn the target source and the system noise ON/OFF at will. If we turn OFF the target source but keep the system noise ON and observe the response of the system in the target direction, we will see random noise that can be quantified by its standard deviation; this is the ``system equivalent'' component. If we now reverse the situation by turning OFF the system noise and turning ON the target source, the \text{SEFD} is that flux density for which the \emph{expected value} of the measurement due to the target source is equal to the \emph{standard deviation} due to the system noise.  Note that \emph{no assumption} is made regarding the polarization state of the target source. Also because the target is turned OFF during the system noise measurement, it stands to reason that the resulting \text{SEFD} is \emph{independent} of the target electric field. Hence, we expect a sensitivity figure of merit that characterizes only the system, which is a desirable outcome. The conceptual approach outlined above, coupled with the statistical independence of the noise sources (as will be explained later), results in an efficient calculation of sensitivity in terms of SEFD. This result allows us to derive a measure with units of \si{\metre\squared\per\kelvin} which is comparable to $A_e/T_{sys}$ but applicable to the polarimeter.

\section{Theory and SEFD calculation}
\label{sec:thy}
\subsection{The Response to the Target Field}
\label{sec:Etarget}
The open-circuit voltages seen at the inputs of the amplifiers connected to the antenna system 1 due to the target electric field ($|_t$) is~\citep{Smirnov1_2011, doi:10.1002/2014RS005517, sokolowski_colegate_sutinjo_2017}\footnote{In this paper and as shown in Fig.~\ref{fig:dual_pol}, we use the standard spherical coordinate system~\citep[see][chap.~2]{Kraus} and \cite{6758443}. In this convention, the $X$ antenna is aligned with the $x$-axis and the $Y$ antenna with the $y$-axis at each dual-polarized antenna location. The angle $\phi$ is the angle with respect to the x-axis and $\theta$ to the $z$-axis. The polarization unit vectors $\hat{\phi}$ and $\hat{\theta}$ are tangential to the circles of constant $\theta$ and constant $\phi$, respectively. The vector directions are determined by the directions of increasing $\phi$ and $\theta$, respectively. Therefore at zenith $(\theta=\SI{0}{\degree},\phi=\SI{0}{\degree})$, the co-polarized component of the X antenna is $\hat{\theta}$ and the co-polarized component of the Y antenna is $\hat{\phi}$. Hence, $l_{X\theta}$ and $l_{Y\phi}$ are the diagonal entries in the Jones matrix in equation~\ref{eqn:J1}. }
\begin{eqnarray}
\mathbf{v}_1|_t&=&\mathbf{J}\mathbf{e}_t, \nonumber \\
\left[ \begin{array}{c}
V_{X1}|_t \\
V_{Y1}|_t 
\end{array} \right]
&=&
\left[ \begin{array}{cc}
l_{X\theta} & l_{X\phi} \\
l_{Y\theta} & l_{Y\phi}
\end{array} \right]
\left[ \begin{array}{c}
E_{t\theta} \\
E_{t\phi} 
\end{array} \right].
\label{eqn:J1}
\end{eqnarray}
The entries of the Jones matrix $\mathbf{J} $ has a unit of antenna length in $\si{\metre}$ which transforms the electric field in \si{\volt\per\metre} to voltage. The Jones matrix entries are direction dependent and deterministic quantities. The electric field components $E_{t\theta}, E_{t\phi}$ as well as the port voltages are complex random variables which we treat as complex envelope quantities. Similarly, the response of antenna system 2 to the same target is
\begin{eqnarray}
\mathbf{v}_2|_t=e^{j\frac{2\pi}{\lambda}f(\bar{r}_{12},\theta_t,\phi_t)}\mathbf{J}\mathbf{e}_t. 
\label{eqn:J2}
\end{eqnarray}
We assume that the antenna system 2 consists of  dual-polarized antennas of identical design and response as antenna system 1. The exponential term in equation~\ref{eqn:J2} is a known phase shift due to the direction of the target and the position of antenna system 2 relative to antenna system 1. The exponent term can be removed from the measurement and will no longer be carried henceforth.   

The correlator forms the outer product 
\begin{eqnarray}
\mathbf{v}_1\mathbf{v}_2^H|_t=\mathbf{J}\mathbf{e}_t\mathbf{e}_t^H\mathbf{J}^H=
\left[ \begin{array}{cc}
V_{X1}V_{X2}^*|_t & V_{X1}V_{Y2}^*|_t \\
V_{Y1}V_{X2}^*|_t & V_{Y1}V_{Y2}^*|_t
\end{array} \right].
\label{eqn:outer_voltage}
\end{eqnarray}
The expected value of equation~\ref{eqn:outer_voltage} is
\begin{eqnarray}
\left<\mathbf{v}_1\mathbf{v}_2^H\right>|_t=\mathbf{J}\left<\mathbf{e}_t\mathbf{e}_t^H\right>\mathbf{J}^H,
\label{eqn:exputer_voltage}
\end{eqnarray}
where the quantity of interest is 
\begin{eqnarray}
\left<\mathbf{e}_t\mathbf{e}_t^H\right>&=&
\left[ \begin{array}{cc}
\left<|E_{t\theta}|^2\right>& \left<E_{t\theta}E_{t\phi}^*\right> \\
\left<E_{t\theta}^*E_{t\phi} \right> & \left<|E_{t\phi}|^2\right>
\end{array} \right]\nonumber \\
&=&\frac{1}{2}\left[ \begin{array}{cc}
I+Q & U-jV \\
U+jV & I-Q
\end{array}\right].
\label{eqn:outer_field}
\end{eqnarray}
from which the polarization property of the target is inferred using the Stokes parameters $I, Q, U, V$~\citep{Wilson2009, Smirnov1_2011}\footnote{Equation~\ref{eqn:outer_field} produces $V=1$ (positive) for incident right-hand circularly polarized wave (RCHP). Positive $V$ for RHCP is consistent with \cite{IAU1973}, \cite{1996A&AS..117..161H}, \cite{Smirnov1_2011} and \cite{8657413}. 
Equation~7 in \cite{Smirnov1_2011} has $U+iV$ and $U-iV$ in the top right and bottom left, respectively. Most likely, this is because \cite{1996A&AS..117..161H, Smirnov1_2011} use the International Astronomical Union (IAU) coordinate system (see: https://healpix.jpl.nasa.gov/html/intronode12.htm) whereas we use the spherical coordinate system. Note that the unit vector $\hat{\theta}$ in the spherical coordinate system points ``South'' ($-\hat{X}$) in the IAU coordinate system.}. 

\subsection{SEFD}
\label{sec:SEFD}

In practice we obtain equation~\ref{eqn:outer_field} from $\mathbf{v}_1\mathbf{v}_2^H$ which is dominated by system noise, such that the result is an estimate of equation~\ref{eqn:outer_field}
\begin{eqnarray}
\tilde{\mathbf{e}}\tilde{\mathbf{e}}^H=\mathbf{J}^{-1}\mathbf{v}_1\mathbf{v}_2^H\mathbf{J}^{-H}.
\label{eqn:pol}
\end{eqnarray}
The Jones matrix must be invertible for polarimetry. As seen in Section~\ref{sec:shor_dipole} later, the Jones matrix of a dual-polarized dipole antenna system is invertible for angles above the horizon. The flux density of the target is defined as
\begin{eqnarray}
S_t&=&\frac{\left<\mathbf{e}_t\mathbf{e}_t^H\right>_{1,1}+\left<\mathbf{e}_t\mathbf{e}_t^H\right>_{2,2}}{\eta_0}=\frac{\left<|E_{t\theta}|^2\right>+\left<|E_{t\phi}|^2\right>}{\eta_0}\nonumber \\
&=&\frac{I}{\eta_0},
\label{eqn:St}
\end{eqnarray}
in \si{\watt\per\square\metre} and $\eta_0=\sqrt{\mu_0/\epsilon_0}\approx\SI{120\pi}{\ohm}$ is the free space impedance; the subscript $\left<.\right>_{\_,\_}$ indicates the matrix entry; the magnitude of the complex envelope is taken as the rms value. The $\text{SEFD}_I$ (strictly speaking, the underscore $._I$ is redundant but we choose to keep it to distinguish this true SEFD from the $\text{SEFD}_{unpol.}$ based on the assumption of an unpolarized target because the latter is so pervasive) therefore is the flux density of a target source which is equal to the standard deviation of the sum of the diagonal of the noisy estimate $\tilde{\mathbf{e}}\tilde{\mathbf{e}}^H$,
\begin{eqnarray}
\text{SEFD}_I=\frac{\text{SDev}\left[(\tilde{\mathbf{e}}\tilde{\mathbf{e}}^H)_{1,1}+(\tilde{\mathbf{e}}\tilde{\mathbf{e}}^H)_{2,2}\right]}{\eta_0}.
\label{eqn:SEFD_STDev}
\end{eqnarray}

The standard deviations of each diagonal entry of $\tilde{\mathbf{e}}\tilde{\mathbf{e}}^H$ are also quantities of interest because these entries are produced in a typical radio astronomy correlation process. These quantities are conventionally called  $XX$ and $YY$ after the diagonals of equation~\ref{eqn:outer_voltage}. However, in the spherical coordinate system, they are $\theta\theta$ and $\phi\phi$ because they are estimates of $|E_{\theta}|^2$ and $|E_{\phi}|^2$. Note that, strictly speaking, we should not attempt to define SEFD for $\theta\theta$ or $\phi\phi$ quantities separately because each one has insufficient information for reconstructing the flux density of an arbitrarily polarized target. 

\subsection{Calculation of standard deviation due to system noise}
\label{sec:STdev_Calc}
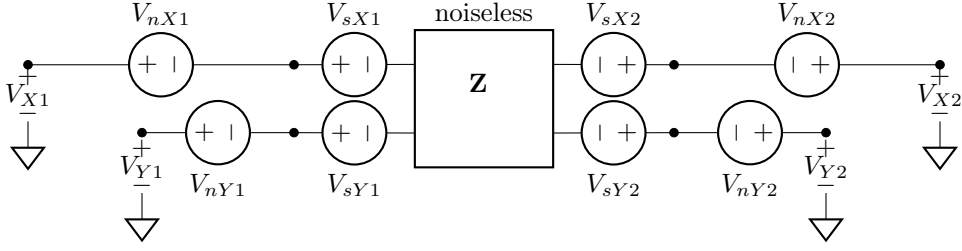
\begin{figure*}
    \centering
\begin{circuitikz}[scale = 1]
\draw 
(0,0) node[fourport] (c) {noiseless}
(c.4) to[short] ++(0,0) to [V,invert,-*,l_=$V_{sX1}$] 
(-2.5,0.45) to [V,invert,-*,l_=$V_{nX1}$] (-6,0.45)
(c.3) to[short] ++(0,0) to [V,invert,-*,l=$V_{sX2}$] (2.5,0.45)
to [V,invert,-*,l=$V_{nX2}$] (6,0.45)
(c.1) to[short] ++(0,0) to [V,invert,-*,l^=$V_{sY1}$] 
(-2.5,-0.45) to [V,invert,-*,l^=$V_{nY1}$] (-4.5,-0.45) 
(c.2) to[short,-] ++(0,0) to [V,invert,-*,l_=$V_{sY2}$] (2.5,-0.45)
to [V,invert,-*,l_=$V_{nY2}$] (4.5,-0.45) 
(c.4) node[below right] {$~~~~~\mathbf{Z}$}
;
\draw
(-6,0.40) [open, v=$V_{X1}$] to (-6,-.3) node[sground] {} 
(-4.5,-0.55) [open, v=$V_{Y1}$] to (-4.5,-1.25) node[sground] {} 
(6,0.40) [open, v^=$V_{X2}$] to (6,-.3) node[sground] {} 
(4.5,-0.55) [open, v^=$V_{Y2}$] to (4.5,-1.25) node[sground] {} 
;
\end{circuitikz}
    \caption{Circuit representation of the four-port network. The impedance matrix $\mathbf{Z}$ is a noiseless network describing relationship between the antenna voltages and currents. The sky noise and the amplifier noise sources are denoted by subscripts $._s$ and $._n$, respectively.}
    \label{fig:antenna_noise}
\end{figure*}

In this section, the target field is turned OFF. The system noise consists of sky noise and amplifier noise represented by voltage sources with subscripts $._s$ and $._n$, respectively in Fig.~\ref{fig:antenna_noise}. For example, $V_{nX1}$ is the voltage noise due to the amplifier (more generally the receiver chain) connected to antenna $X1$; $V_{nY2}$ is the voltage noise due to the amplifier connected to antenna $Y1$;  $V_{sX1}$ and $V_{sY2}$ represent the sky noise seen at the antennas $X1$ and $Y2$, respectively; $V_{X1}$ and $V_{Y2}$ represent the total voltages seen at antennas $X1$ and $Y2$, respectively, which are the quantities that are processed by the correlator. The interaction between voltage and current at the antenna ports is captured by the impedance matrix $\mathbf{Z}$~\citep{doi:10.1002/2014RS005517, warnick_maaskant_ivashina_davidson_jeffs_2018}. The noise sources due to the sky and the amplifiers are mutually independent. Every antenna is connected to its own receiver chain such that the receiver noise is independent from one another. The representation in Fig.~\ref{fig:antenna_noise} is a standard treatment for noise in a multiport system with a well-established history~\citep{doi:10.1063/1.1722048, 4052096, doi:10.1002/jnm.1660030408} and application in receiving phased array systems~\citep{Warnick_5062509, warnick_maaskant_ivashina_davidson_jeffs_2018}.

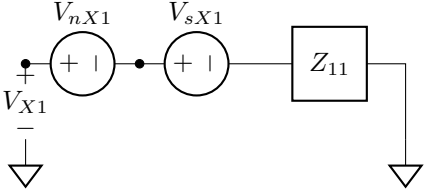
\begin{figure}[t]
    \centering
\begin{circuitikz}[scale = 1]
\draw 
(0,0) to[twoport,t=$Z_{11}$] ++(2,0) to[short] (2,-1) node[sground]{}
(0,0) to [V,invert,-*,l_=$V_{sX1}$] (-1.5,0) 
to [V,invert,-*,l_=$V_{nX1}$] (-3,0) 
(-3,0) [open, v=$V_{X1}$] to (-3,-1) node[sground] {} ;
\end{circuitikz}
    \caption{The simplified impedance (noiseless) and noise sources seen at the antenna port. $Z_{11}$ is the impedance seen at the antenna port 1. The same circuit is seen at $Y1, X2, Y2$ with subscript substitutions.}
    \label{fig:indep_noise}
\end{figure}

In the absence of the target field, the interferometer is illuminated by a diffuse sky noise such that the mutual coherence between antenna systems 1 and 2 is insignificant for separations greater than tens of wavelengths~\citep{7293140}. In addition, for orthogonally polarized antenna elements, the isolation between the two polarizations (e.g. $X1$ and $Y1$) is high such that the mutual coherence between them due to the diffuse sky is negligible. In antenna engineering terms, the antenna mutual coupling is considered insignificant for statistical calculation. This permits all voltage noise sources in Fig.~\ref{fig:antenna_noise} to be treated as independent noise which greatly simplifies the computation. Furthermore, if we assume all antennas are of identical design, the impedance seen into each antenna port ($Z_{11}=Z_{ant}$) is identical at all four ports as shown in Fig.~\ref{fig:indep_noise}.  

We base our standard deviation calculation on the simplified circuit in Fig.~\ref{fig:indep_noise}. For that we need the statistics of $V_n$ and $V_s$.
\begin{itemize}
    \item $V_s$ is a thermal noise source due to antenna resistance $R_{ant}$ (the real part of $Z_{ant}$) at antenna temperature $T_{ant}$  which is the result of the integral of the product of the antenna pattern  and $T_{sky}$~\citep[see][chap.~17]{Kraus}.  
    \item $V_n$ is a also thermal noise source due to the amplifier noise which is represented by the same antenna resistance $R_{ant}$ at receiver noise temperature $T_{rx}$. 
    \item All voltage noise sources are mutually independent.
    \item We can combine $T_{sys}=T_{ant}+T_{rx}$ as the system noise temperature of $R_{ant}$.
\end{itemize}

The correlator output due to the sky and amplifier noise sources is 
\begin{eqnarray}
\mathbf{v}_1\mathbf{v}_2^H|_{n,s}=
\left[ \begin{array}{cc}
V_{X1}V_{X2}^*|_{n,s} & V_{X1}V_{Y2}^*|_{n,s} \\
V_{Y1}V_{X2}^*|_{n,s} & V_{Y1}V_{Y2}^*|_{n,s}
\end{array} \right],
\label{eqn:outer_voltage_n_s}
\end{eqnarray}
where $V_{X1}=V_{sX1}+V_{nX1}$ and similarly for all other ports. The noisy estimates of the electric field outer product are formed by flanking equation~\ref{eqn:outer_voltage_n_s} with the Jones matrix inverse as shown in equation~\ref{eqn:pol}. For brevity, we drop the leading $V$ and introduce shorthand notation, e.g. $V_{X1}V_{X2}^*|_{n,s}=X_1X_2^*$ etc. This results in
\begin{eqnarray}
|D|^2(\tilde{\mathbf{e}}\tilde{\mathbf{e}}^H)_{1,1}&=&|l_{Y\phi}|^2X_1X_2^*-l_{Y\phi}l_{X\phi}^*X_1Y_2^* \nonumber \\ &-&l_{X\phi}l_{Y\phi}^*Y_1X_2^*+|l_{X\phi}|^2Y_1Y_2^*, \nonumber \\
|D|^2(\tilde{\mathbf{e}}\tilde{\mathbf{e}}^H)_{2,2}&=&
|l_{Y\theta}|^2X_1X_2^*-l_{Y\theta}l_{X\theta}^*X_1Y_2^* \nonumber \\
&-&l_{X\theta}l_{Y\theta}^*Y_1X_2^*+|l_{X\theta}|^2Y_1Y_2^*,
\label{eqn:eeh11_22}
\end{eqnarray}
where
\begin{eqnarray}
D=\det \mathbf{J}=l_{X\theta}l_{Y\phi}-l_{X\phi}l_{Y\theta}
\label{eqn:det_J}
\end{eqnarray}
is the determinant of the Jones matrix. 

\subsubsection{Standard deviations of $\theta\theta$ and $\phi\phi$}
\label{sec:SEFD_tht_phi}
It can be shown (see Appendix)
\begin{eqnarray}
|D|^4\text{Var}(\tilde{\mathbf{e}}\tilde{\mathbf{e}}^H)_{1,1}&=&|l_{Y\phi}|^4\left<|X_1|^2\right>\left<|X_2|^2\right>\nonumber \\
&+&|l_{Y\phi}|^2|l_{X\phi}|^2\left<|X_1|^2\right>\left<|Y_2|^2\right> \nonumber \\ &+&|l_{X\phi}|^2|l_{Y\phi}|^2\left<|Y_1|^2\right>\left<|X_2|^2\right>\nonumber\\
&+&|l_{X\phi}|^4\left<|Y_1|^2\right>\left<|Y_2|^2\right>. 
\label{eqn:var_11}
\end{eqnarray}
Because the antennas are of identical design, the mean-square noise voltages are $\left<|X_1|^2\right>=\left<|X_2|^2\right>=4kT_{sysX}R_{ant}\Delta f$ and $\left<|Y_1|^2\right>=\left<|Y_2|^2\right>=4kT_{sysY}R_{ant}\Delta f$ with units \si{\volt\squared}, which are expected for a resistance $R_{ant}$ at temperature $T_{sys}$ with noise bandwidth $\Delta f$. As a result, we get
\begin{eqnarray}
\frac{\text{Var}(\tilde{\mathbf{e}}\tilde{\mathbf{e}}^H)_{1,1}}{{(4kR_{ant}\Delta f)^2}}=\frac{(|l_{Y\phi}|^2T_{sysX} +|l_{X\phi}|^2T_{sysY})^2}{|D|^4},
\label{eqn:var_11_simp}
\end{eqnarray}
and taking the square root
\begin{eqnarray}
\frac{\text{SDev}(\tilde{\mathbf{e}}\tilde{\mathbf{e}}^H)_{1,1}}{4kR_{ant}\Delta f}=\frac{|l_{Y\phi}|^2T_{sysX} +|l_{X\phi}|^2T_{sysY}}{|D|^2}.
\label{eqn:sdev_11_simp}
\end{eqnarray}
Similarly,
\begin{eqnarray}
\frac{\text{SDev}(\tilde{\mathbf{e}}\tilde{\mathbf{e}}^H)_{2,2}}{4kR_{ant}\Delta f}=\frac{|l_{Y\theta}|^2T_{sysX} +|l_{X\theta}|^2T_{sysY}}{|D|^2}.
\label{eqn:sdev_22_simp}
\end{eqnarray}

\subsubsection{Calculation of $\text{SEFD}_{I}$}
\label{sec:SEFD_I_calc}
Summing the two equations in equation~\ref{eqn:eeh11_22}, we obtain
\begin{eqnarray}
|D|^2\tilde{I}&=&X_1X_2^*[|l_{Y\phi}|^2+|l_{Y\theta}|^2] \nonumber\\ 
&+&Y_1Y_2^*[|l_{X\phi}|^2+|l_{X\theta}|^2] \nonumber\\
&-&X_1Y_2^*[l_{X\phi}^*l_{Y\phi}+l_{X\theta}^*l_{Y\theta}] \nonumber\\
&-&Y_1X_2^*[l_{X\phi}l_{Y\phi}^*+l_{X\theta}l_{Y\theta}^*], 
\label{eqn:Isimp}
\end{eqnarray}
where $\tilde{I}=(\tilde{\mathbf{e}}\tilde{\mathbf{e}}^H)_{1,1}+(\tilde{\mathbf{e}}\tilde{\mathbf{e}}^H)_{2,2}$. Following the same reasoning in Sec.~\ref{sec:SEFD_tht_phi} 
\begin{eqnarray}
\frac{|D|^4\text{Var}(\tilde{I})}{{(4kR_{ant}\Delta   f)^2}}&=&\norm{\mathbf{l}_Y}^4T_{sysX}^2  
+\norm{\mathbf{l}_X}^4T_{sysY}^2  \nonumber \\
&+&2|l_{X\phi}^*l_{Y\phi}+l_{X\theta}^*l_{Y\theta}|^2T_{sysX}T_{sysY}.
\label{eqn:var_I_simp}
\end{eqnarray}
where the vector norms are
\begin{eqnarray}
\norm{\mathbf{l}_X}^2&=&|l_{X\theta}|^2+|l_{X\phi}|^2, \nonumber \\
\norm{\mathbf{l}_Y}^2&=&|l_{Y\theta}|^2+|l_{Y\phi}|^2. 
\label{eqn:normL}
\end{eqnarray}
Taking $\sqrt{\text{Var}(\tilde{I})}$ from equation~\ref{eqn:var_I_simp} and dividing by $\eta_0$ produces the desired result
\begin{eqnarray}
\text{SEFD}_{I}=\frac{\text{SDev}(\tilde{I})}{\eta_0}=\Delta fk\frac{4 R_{ant}}{\eta_0}\frac{\sqrt{L_T}}{|D|^2},
\label{eqn:SEFD_I_res}
\end{eqnarray}
with units \si{\watt\per\metre\squared}. If $\mathrm{SEFD}_I$ is stated in \si{\watt\per\metre\squared\per\hertz}, which is the case for noise-like sources in radio astronomy, then we remove $\Delta f$ from the right hand side. The $\Delta f$ provides the flexibility to account for deterministic sources  (e.g., a continuous wave with signal bandwidth less than $\Delta f$) of unknown polarization if required. $D$ is the Jones matrix determinant in equation~\ref{eqn:det_J} and
\begin{eqnarray}
L_T&=&\norm{\mathbf{l}_Y}^4T_{sysX}^2 +\norm{\mathbf{l}_X}^4T_{sysY}^2 \nonumber\\
&+&2|l_{X\phi}^*l_{Y\phi}+l_{X\theta}^*l_{Y\theta}|^2T_{sysX}T_{sysY};
\label{eqn:L_T}
\end{eqnarray}
$\sqrt{L_T}/|D|^2$ has units of \si{\kelvin\per\metre\squared}. 

\section{SEFD interpretation, special cases, examples and application}
\label{sec:examples}
Equation~\ref{eqn:SEFD_I_res} is the desired result, but it bears no resemblance to commonly used quantities in radio astronomy which warrants further explanation. The purpose of this section is to interpret equation~\ref{eqn:SEFD_I_res} by linking it to more familiar quantities, which is possible under certain assumptions and approximation.  

\subsection{Comparable measure to $A/T$}
\label{sec:AonT}
It is generally not possible to factor out $A_e$ and $T_{sys}$ as separate entities from equation~\ref{eqn:SEFD_I_res}. This is because $T_{sysX}$ and  $T_{sysY}$ are inextricably linked with the components of the Jones matrix as shown in equation~\ref{eqn:L_T}. Still, because of the long tradition of using of $A_e/T_{sys}$, it is useful to produce a comparable metric that has units of \si{\metre\squared\per\kelvin}. This can be done by simply inverting equation~\ref{eqn:SEFD_I_res} and removing the Boltzmann constant, $k$,
\begin{eqnarray}
\mathrm{AonT}_I=\frac{\eta_0}{4 R_{ant}}\frac{|D|^2}{\sqrt{L_T}}.
\label{eqn:AoT}
\end{eqnarray}

\subsubsection{Special case $A/T$ for $T_{sysX}=T_{sysY}=T_{sys}$}
\label{sec:AonT_same_tsys} 
In this condition, $T_{sys}$ factors out of $\sqrt{L_T}$ and
we can write
\begin{eqnarray}
\mathrm{AonT}_I|_{T_{sysX}=T_{sysY}}=\frac{\eta_0}{4 R_{ant}}\frac{|D|^2}{T_{sys}\sqrt{L}},
\label{eqn:AoT_eq_Tsys}
\end{eqnarray}
where
\begin{eqnarray}
L=\norm{\mathbf{l}_Y}^4 +\norm{\mathbf{l}_X}^4 
+2|l_{X\phi}^*l_{Y\phi}+l_{X\theta}^*l_{Y\theta}|^2.
\label{eqn:L}
\end{eqnarray}

\subsection{Short dipoles}
\label{sec:shor_dipole}
Let the antenna elements be orthogonal crossed short dipoles with $X$ dipole located on the X-Z plane ($\phi=0\degree$) and $Y$ dipole located on the Y-Z plane ($\phi=90\degree$)~\citep{doi:10.1002/2014RS005517}
\begin{eqnarray}
\mathbf{J}
=l_a(\theta)\left[ \begin{array}{cc}
\cos\theta\cos\phi & -\sin\phi \\
\cos\theta\sin\phi & \cos\phi
\end{array} \right];
\label{eqn:Jdipole}
\end{eqnarray}
let $l_a(\theta)$ be a scalar multiplier that represents the electrical length of the antenna and the array factor due to a conductive ground screen. In this case
\begin{eqnarray}
|D|^2=l_a(\theta)^4\cos^2\theta,
\label{eqn:D2_dip}
\end{eqnarray}
and 
\begin{eqnarray}
L=l_a(\theta)^4(1+\cos^4\theta), 
\label{eqn:L_dip}
\end{eqnarray}
which are independent of $\phi$. For $T_{sysX}=T_{sysY}=T_{sys}$,
\begin{eqnarray}
\mathrm{AonT}_I|_{T_{sysX}=T_{sysY}}=\frac{\eta_0}{4 R_{ant}}\frac{l_a(\theta)^2}{T_{sys}\sqrt{\frac{1}{\cos^4\theta}+1}}.
\label{eqn:AoT_eq_Tsys_short_dip}
\end{eqnarray}
\subsubsection{Connection to Jones matrix singular values and condition number of a short dipole}
\label{sec:conIXR}
The $\phi$ independence of equation~\ref{eqn:AoT_eq_Tsys_short_dip} and its dependence on $\cos^4\theta$ merits further discussion because they are connected to the singular values of the Jones matrix of a short dipole. It has been shown in~\cite{8878981} that the ratio of the max./min. of the singular values (i.e., the condition number, $c\{\mathbf{J}\}$) of the Jones matrix determines the intrinsic cross-polarization ratio, IXR=$(c\{\mathbf{J}\}+1)^2/(c\{\mathbf{J}\}-1)^2$, of the dual-polarized antenna system. The Jones matrix of the orthogonal short dipole system in~equation~\ref{eqn:Jdipole} is expressed as singular value decomposition
\begin{eqnarray}
\mathbf{J}&=&\left[ \begin{array}{cc}
\cos\phi & -\sin\phi \\
\sin\phi & \cos\phi
\end{array} \right]
\left[ \begin{array}{cc}
l(\theta) \cos\theta  & 0 \\
0 & l(\theta)
\end{array} \right] \mathbf{I} \nonumber \\
&=&\mathbf{U}\mathbf{\Sigma}\mathbf{V}^H.
\label{eqn:Jdipole_SVD}
\end{eqnarray}
where $\mathbf{U}$ is an orthogonal rotation matrix and $\mathbf{\Sigma}$ is a diagonal matrix of singular values $\sigma_{max}=l(\theta)$ and $\sigma_{min}=l(\theta)\cos\theta$. The condition number is 
\begin{eqnarray}
c\{\mathbf{J}\}=\frac{\sigma_{max}}{\sigma_{min}}=\frac{1}{\cos\theta}.
\label{eqn:Jdipole_sigma}
\end{eqnarray}
Hence equation~\ref{eqn:AoT_eq_Tsys_short_dip} may be written as
\begin{eqnarray}
\mathrm{AonT}_I|_{T_{sysX}=T_{sysY}}=\frac{\eta_0}{4 R_{ant}}\frac{l_a(\theta)^2}{T_{sys}\sqrt{c\{\mathbf{J}(\theta)\}^4+1}}.
\label{eqn:AoT_eq_Tsys_short_dip_cond}
\end{eqnarray}

\subsection{Cardinal planes of orthogonal dual-polarized linear antennas}
\label{sec:dual-pol_lin}
For ideal dual-polarized linear antennas, the Jones matrix becomes diagonal or antidiagonal in the cardinal planes. For example at $\phi=0\degree$,  equation~\ref{eqn:Jdipole} becomes diagonal. In practice, this is \emph{approximately} true. In this special case $|D|^2=|l_{X\theta}|^2|l_{Y\phi}|^2$ and
\begin{eqnarray}
L_T(\phi=0\degree)=|l_{Y\phi}|^4T_{sysX}^2 +|l_{X\theta}|^4T_{sysY}^2, 
\label{eqn:L_T_phi0}
\end{eqnarray}
which leads to
\begin{eqnarray}
\text{SEFD}_{I}(\phi=0\degree)
=
k\frac{4 R_{ant}}{\eta_0}\sqrt{\frac{T_{sysX}^2}{|l_{X\theta}|^4}+\frac{T_{sysY}^2}{|l_{Y\phi}|^4}},
\label{eqn:SEFD_I_res_phi0}
\end{eqnarray}
where the SEFD is given in \si{\watt\per\metre\squared\per\hertz} and is valid for arbitrary polarization. We see in equation~\ref{eqn:SEFD_I_res_phi0} the sum of the squares of $X$-only and $Y$-only quantities under the square root. This is because the $\theta\theta$  and $\phi\phi$ become fully independent measurements on the cardinal plane. Equations \ref{eqn:sdev_11_simp} and \ref{eqn:sdev_22_simp} become
\begin{eqnarray}
\text{SDev}(\tilde{\mathbf{e}}\tilde{\mathbf{e}}^H)_{1,1}(\phi=0\degree)&=& 4kR_{ant}\Delta f\frac{T_{sysX}}{|l_{X\theta}|^2}, \nonumber\\
\text{SDev}(\tilde{\mathbf{e}}\tilde{\mathbf{e}}^H)_{2,2}(\phi=0\degree)&=&4kR_{ant}\Delta f\frac{T_{sysY}}{|l_{Y\phi}|^2},
\label{eqn:Sdev_theta_phi}
\end{eqnarray}
and the standard deviation of the sum becomes the square root of the sum of the squares as expected. For this special case, we can make a connection to the $\text{SEFD}_{unpol.}$ \emph{assumption} as described in equation~\ref{eqn:SEFD_AonT_formula}. Following equation~\ref{eqn:sdev_11_simp} and equation~\ref{eqn:sdev_22_simp} we can write
\begin{eqnarray}
\text{SEFD}_{XX}(\phi=0\degree)&=&2k \frac{4R_{ant}}{\eta_0}\frac{T_{sysX}}{|l_{X\theta}|^2}, \nonumber\\
\text{SEFD}_{YY}(\phi=0\degree)&=&2k \frac{4R_{ant}}{\eta_0}\frac{T_{sysY}}{|l_{Y\phi}|^2},
\label{eqn:SEFD_XXYY}
\end{eqnarray}
such that
\begin{eqnarray}
\text{SEFD}_{I}\approx\frac{1}{2}\sqrt{\text{SEFD}_{XX}^2+\text{SEFD}_{YY}^2},~\text{for}~\phi=0\degree,90\degree. 
\label{eqn:SEFD_I_phi0_in_XXYY}
\end{eqnarray}
The $\approx$ sign serves as a reminder that the diagonal/anti-diagonal Jones matrix may be approached in practice but rarely fulfilled perfectly. It should be noted that equation~\ref{eqn:SEFD_I_phi0_in_XXYY} is valid for the cardinal plane only but is \emph{independent} of the polarization state of the target field \emph{even though} $\text{SEFD}_{XX}$ and $\text{SEFD}_{YY}$ assume the polarization mismatch factor 1/2 which is associated with an unpolarized target source. Also, because of the squares, the higher of the $\text{SEFD}_{XX}$ and $\text{SEFD}_{YY}$ will dominate the system $\text{SEFD}_I$. 

\subsection{Limitation of the narrow FoV SEFD approximation}
\label{sec:limitation}
Equation \ref{eqn:SEFD_I_phi0_in_XXYY} has a clear link to ~\cite{1999ASPC..180..171W} in that this is exactly expected for a narrow FoV in the vicinity of the beam center since the FoV is always close to the cardinal planes. What is not expected is that equation~\ref{eqn:SEFD_I_phi0_in_XXYY} is generally incorrect, which becomes evident only when the FoV is very wide. For example, take the short dipole system assuming $T_{sysX}=T_{sysY}=T_{sys}$,
\begin{eqnarray}
\text{SEFD}_{I}=k\frac{4 R_{ant}}{\eta_0}\frac{T_{sys}\sqrt{L}}{|D|^2}=k\frac{4 R_{ant}}{\eta_0}\frac{T_{sys}\sqrt{\frac{1}{\cos^4\theta}+1}}{l_a(\theta)^2}.
\label{eqn:SEFD_I_dip}
\end{eqnarray}
Assuming an unpolarized source, 
\begin{eqnarray}
\text{SEFD}_{XX}&=&2k \frac{4R_{ant}}{\eta_0}\frac{T_{sys}}{\norm{\mathbf{l}_{X}}^2}, \nonumber\\
\text{SEFD}_{YY}&=&2k \frac{4R_{ant}}{\eta_0}\frac{T_{sys}}{\norm{\mathbf{l}_{Y}}^2},
\label{eqn:SEFD_XXYY_dip}
\end{eqnarray}
where 
\begin{eqnarray}
\norm{\mathbf{l}_Y}^2
&=&l_a(\theta)^2(1-\sin^2\theta\sin^2\phi) ,\nonumber \\
\norm{\mathbf{l}_X}^2&=& l_a(\theta)^2(1-\sin^2\theta\cos^2\phi).
\label{eqn:L_dip_parts}
\end{eqnarray}

\begin{figure}[t]
\centering
\noindent
  \includegraphics[width=3.25in]{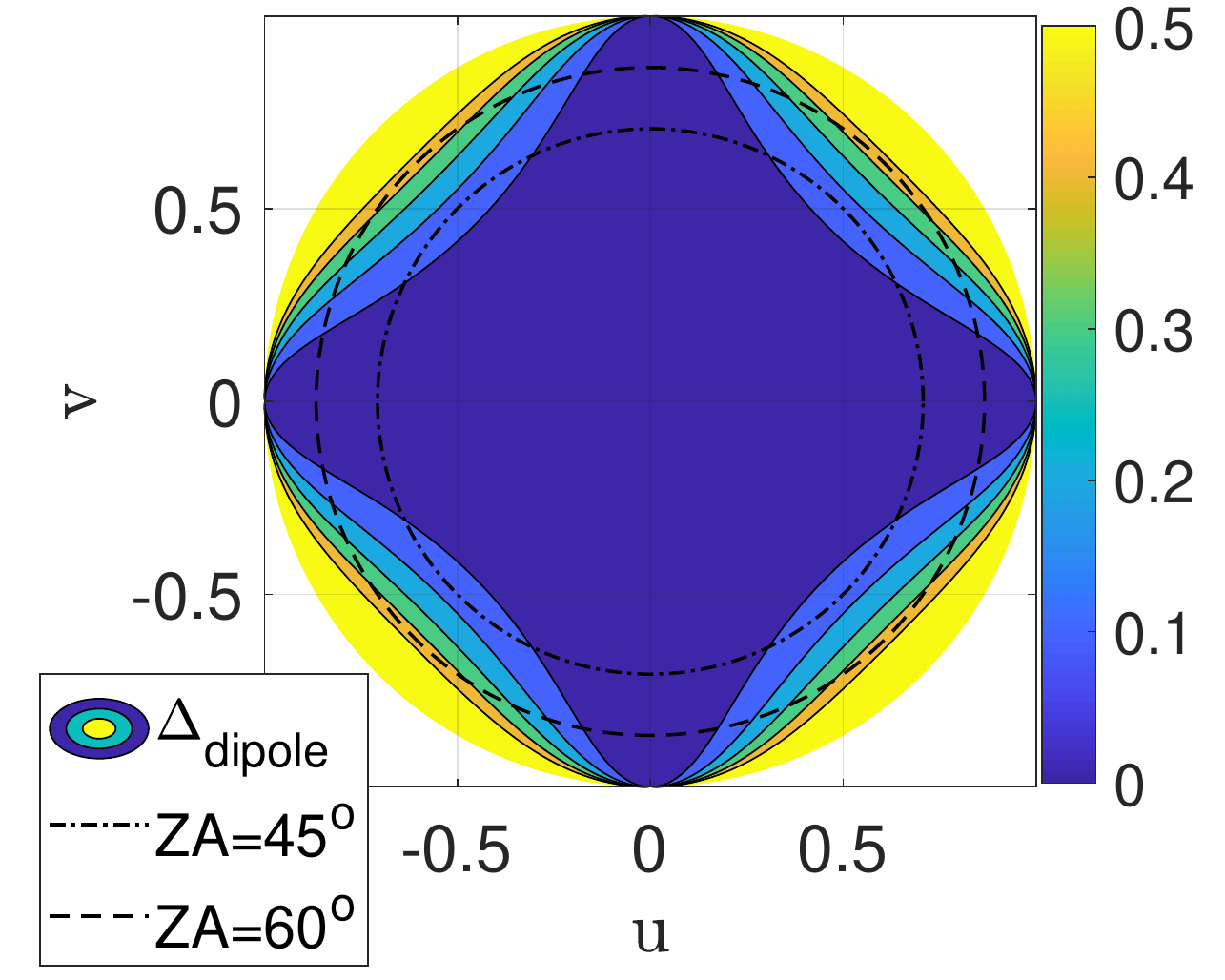}
\caption{The relative error of the narrow FoV SEFD approximation for the dual-polarized short dipole system. $u=\sin\theta\cos\phi$ and $v=\sin\theta\sin\phi$. The ZA circles indicate the zenith angles of observation.}
\label{fig:Hz_dip_delta}
\end{figure}

Fig.~\ref{fig:Hz_dip_delta} shows the relative error of equation~\ref{eqn:SEFD_I_phi0_in_XXYY} if we apply it over the entire sky for the dual-polarized short dipole system
\begin{eqnarray}
\Delta_{\text{dipole}}=\frac{\text{SEFD}_{I}- \frac{1}{2}\sqrt{\text{SEFD}_{XX}^2+\text{SEFD}_{YY}^2}}{\text{SEFD}_{I}}.
\label{eqn:SEFD_I_error}
\end{eqnarray}
It is evident that the maximum relative errors occur at the diagonal planes. For $\text{ZA}=45\degree, 60\degree$ the maximum relative errors are approximately 15\% and 45\%, respectively. Note the error is positive in this example, i.e.   $\text{SEFD}_{I}$ is higher than $\frac{1}{2}\sqrt{\text{SEFD}_{XX}^2+\text{SEFD}_{YY}^2}$.

\subsection{Relevance to next-generation low-frequency radio telescopes, e.g. SKA-Low}
\label{sec:SKA1Low}
Equations \ref{eqn:SEFD_I_res} and \ref{eqn:AoT} have an immediate relevance to SKA-Low sensitivity requirements and interpretation thereof, in particular 
SKA1-SYS\_REQ2135 ``SKA1\_Low array sensitivity'' and SKA1-SYS\_REQ2622 ``Sensitivity for off zenith angles''  in \cite{SKA1L1req}, which is the most recent at the time of writing. REQ2135 calls for  A/T value at zenith ``per polarization'' which implies the assumption of unpolarized source. At zenith, the A/T values of the orthogonal linearly polarized antennas are approximately the same, hence there is no ambiguity. 

REQ2622 calls for maximum allowable degradation (30\% at $60^\circ$ elevation angle; 50\% at $45^\circ$ elevation angle) relative to the peak sensitivity (reasonably assumed to occur at zenith value). However at off-zenith angles, the A/T values of the orthogonally polarized antennas are different, and hence it is ambiguous how the degradation should be computed. Equation \ref{eqn:SEFD_I_res} or \ref{eqn:AoT} removes this ambiguity by providing one sensitivity number over the entire visible sky. Furthermore, the SEFD or AonT given here makes no assumption regarding the polarization state of the source, which removes another limitation.

\begin{figure}[t]
\centering
\noindent
  \includegraphics[width=2.5in]{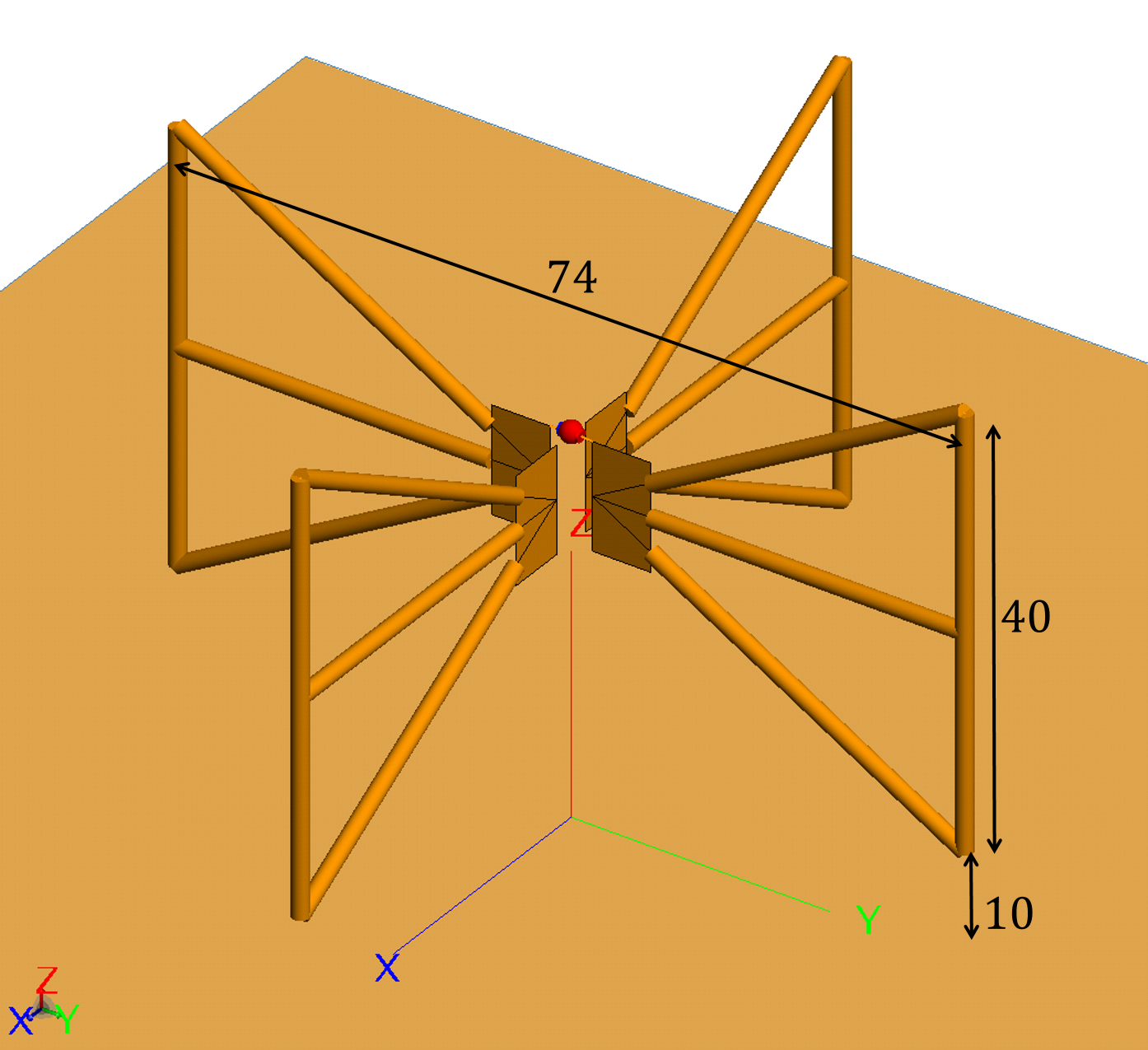}
\caption{The dual-polarized MWA bow-tie antenna simulated in FEKO  (full-wave method-of-moments electromagnetic solver, https://altairhyperworks.com/product/FEKO). The dimensions shown are in \si{\centi\metre}.}
\label{fig:MWA_dip}
\end{figure}

\begin{figure}[t]
\centering
\noindent
  \includegraphics[width=3.25in]{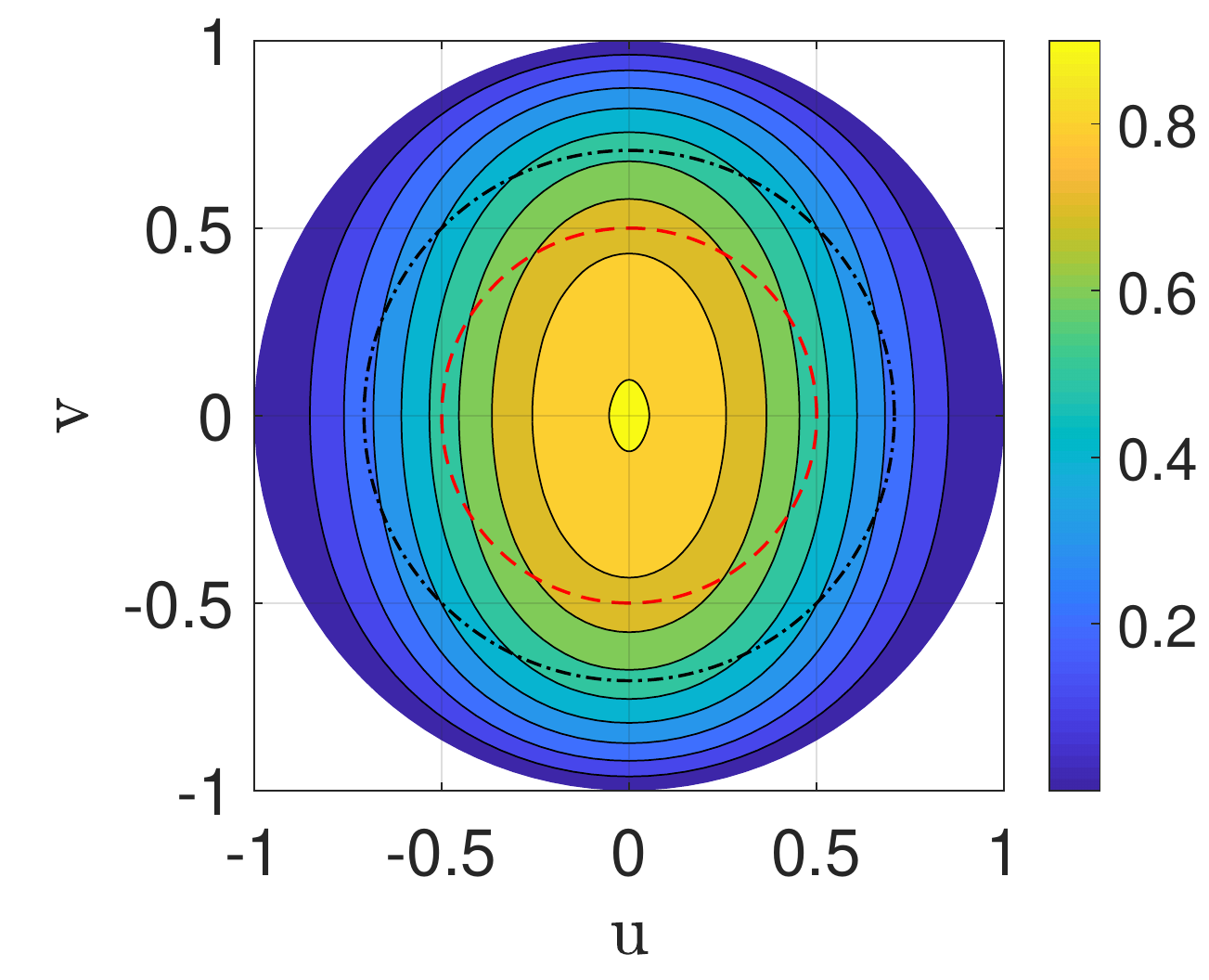}
\caption{Calculated $A_{eX}|_{un.}$ in \si{\metre\squared} at \SI{160}{\mega\hertz} for the MWA bow-tie antenna where $u=\sin\theta\cos\phi$ and $v=\sin\theta\sin\phi$. The red dashed line corresponds to $60^\circ$ elevation angle and the black dash-dot line indicates $45^\circ$ elevation angle.}
\label{fig:MWA_dip_halfAX}
\end{figure}

For example, consider a dual-polarized bow-tie antenna used by the Murchison Widefield Array~\citep[MWA;][]{2013PASA...30....7T, doi:10.1002/2014RS005517, sokolowski_colegate_sutinjo_2017} depicted in Fig.~\ref{fig:MWA_dip}. Suppose we seek to determine its off-zenith performance as suggested in REQ2622. For simplicity let the $T_{sys}$ be equal; hence it is adequate to compare the off-zenith effective area relative to the peak value. Fig.~\ref{fig:MWA_dip_halfAX} shows the simulated effective area assuming an unpolarized source ($A_{eX}|_{un.}$ with factor 1/2 included) for the $X$-directed bow-tie element at \SI{160}{\mega\hertz}. The result for the $Y$-directed element is identical, except for a \SI{90}{\degree} rotation (not shown). It is clearly evident that the effective area defined in this way produces two very different off-zenith values for the $X$ and $Y$ elements, in particular on the cardinal planes. The peak at zenith is \SI{0.9}{\metre\squared} for both $X$ and $Y$ elements, whereas at $\theta=\SI{30}{\degree}, \phi=\SI{0}{\degree}$ $A_{eX}|_{un.}=\SI{0.543}{\metre\squared}$ and $A_{eY}|_{un.}=\SI{0.76}{\metre\squared}$. This results in an ambiguity as to which value (or combination of values) should be used to compute the degradation relative to peak.

\begin{figure}[t]
\centering
\noindent
  \includegraphics[width=3.25in]{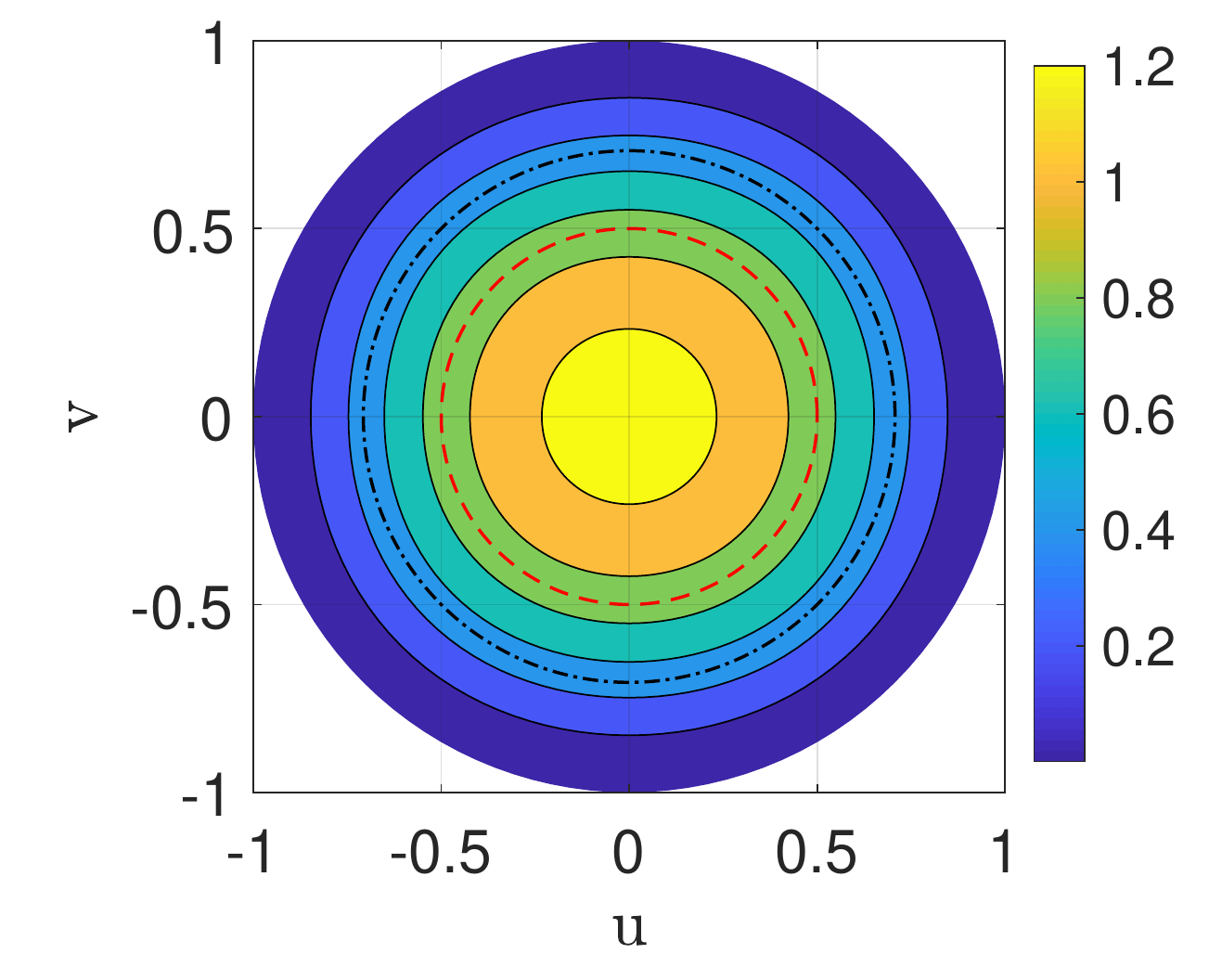}
\caption{Calculated $A_I$ in \si{\metre\squared} at \SI{160}{\mega\hertz} for the MWA bow-tie antenna where $u=\sin\theta\cos\phi$ and $v=\sin\theta\sin\phi$. The red dashed line corresponds to $60^\circ$ elevation angle and the black dash-dot line indicates $45^\circ$ elevation angle.}
\label{fig:MWA_dip_AI}
\end{figure}

Now contrast this with the approach given in this paper. Fig.~\ref{fig:MWA_dip_AI} shows the calculated quantity that we can think of as the  effective area of the \emph{dual-polarized} antenna as per equation~\ref{eqn:AoT_eq_Tsys} (assuming equal $T_{sys}$)
\begin{eqnarray}
A_I=\frac{\eta_0}{4 R_{ant}}\frac{|D|^2}{\sqrt{L}},
\label{eqn:A_I}
\end{eqnarray}
at \SI{160}{\mega\hertz}. The peak value at zenith is \SI{1.28}{\metre\squared}, whereas at $\theta=\SI{30}{\degree}, \phi=\SI{0}{\degree}$ the value is \SI{0.885}{\metre\squared}. The degradation relative to the peak value is simply  $1-0.885/1.28=0.309 \approx 0.3$; there is no ambiguity. 

\section{Validation: all-sky observation and simulation with the MWA}
\label{sec:mwa_obs}

\subsection{Overview and strategy}
\label{sec:valid_overview}
\begin{figure}[h]
\centering
\noindent
  \includegraphics[width=0.3\textwidth]{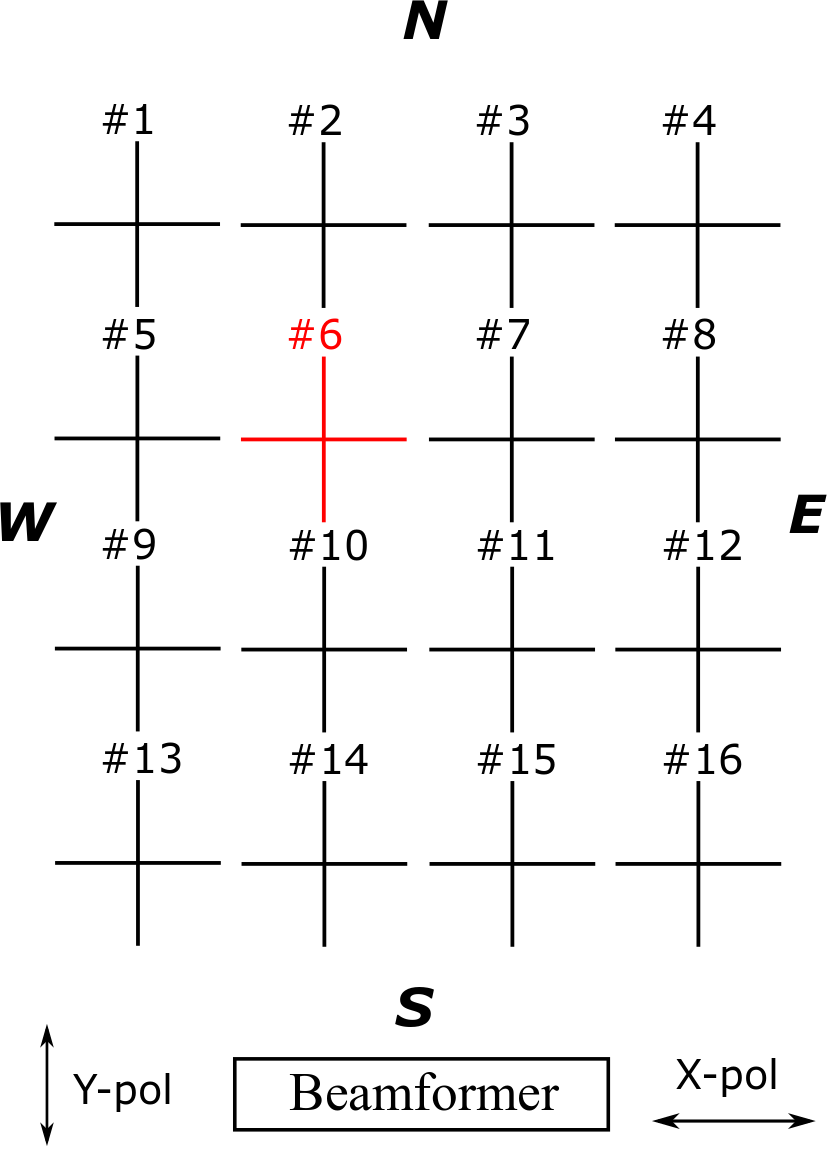}
\caption{An MWA tile of $4\times4$ dual-polarized dipoles. The elements are aligned along N-S and W-E. The spacing between elements is \SI{1.1}{\metre}.}
\label{fig:MWA_tile}
\end{figure}

We use the MWA to validate the proposed $\text{SEFD}_I$ calculation using observational data. The MWA is the SKA-Low precursor operating in the frequency range 70--300 MHz  where an element is a bow-tie, as shown in Fig.~\ref{fig:MWA_dip}. An MWA tile consists of a $4\times4$ array of these antennas. Validation of the equations in this paper is best done using an all-sky observation, as it permits evaluation of SEFD in the regions (off-zenith) where the difference is the most prominent. The array is shown in Fig.~\ref{fig:MWA_tile}, where the element used for all-sky observation (\#6) is highlighted in red. 

\subsection{Simulation predictions for the MWA}\label{sec:mwa_simulations}
\label{sec:mwa_simulation}

The all-sky SEFD was calculated using equation~\ref{eqn:SEFD_I_res} and equation~\ref{eqn:SEFD_I_phi0_in_XXYY} for the MWA tile, with element number 6 being an active element, and the other 15 elements being passive. The excitation amplitude for the active element was "1", while for the passive element, it was "0". Although the passive elements are terminated with the low-noise amplifier (LNA) impedance, they still contribute to the array mutual coupling. Tile beam and scattering parameters were simulated using the full-wave software package FEKO. The frequency of interest is 154.88~MHz, which corresponds to the channel number 121. To calculate the system temperature, we also used measured S-parameters and noise parameters of the LNA \citep{Ung2018}, as well as the Haslam 408 MHz all-sky map \citep{1982A&ASHaslam} scaled to 154.88~MHz using a spectral index of $-2.55$ \citep{spectral_indexASU2019}. The Galactic contribution corresponded to the day and time of the observation. Defined in equation~\ref{eqn:L_T}, $L_T$ is a matrix that requires the system temperatures for both polarizations ($T_{sysX}$, $T_{sysY}$), as well as the effective antenna lengths, which are the values of the Jones matrix in equation~\ref{eqn:J1} computed using the simulated  electric fields $(E_{\theta}, E_{\phi})$ and port currents $I_{X}$, $I_{Y}$ \citep{doi:10.1002/2014RS005517}. For the calculations of $\text{SEFD}_{XX}$ and $\text{SEFD}_{YY}$, the system temperature and antenna effective area were determined according to \cite{Ung2020}. The calculated values are $R_{ant}=\SI{68.4}{\ohm}$, $T_{sysX}=371.04~K$,  $T_{sysY}=348.21~K$.    

\begin{figure}[h]
\centering
\noindent
  \includegraphics[width=0.5\textwidth]{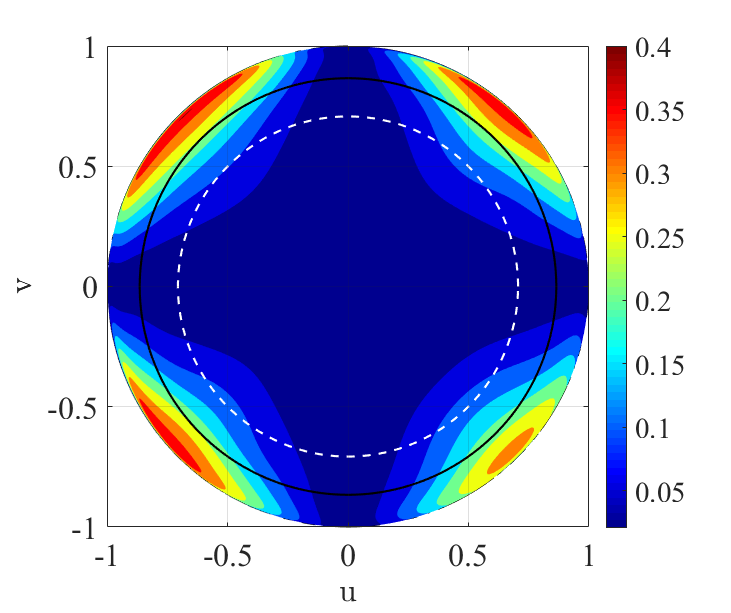}
\caption{The relative difference $\Delta_{\text{dipole}}$ as defined in equation~\ref{eqn:SEFD_I_error} using simulated MWA data. The black solid line is at ZA=60~degrees, while the white dashed line is at ZA=45~degrees.}
\label{fig_mean_ratio_simul}
\end{figure}

The relative error as per equation~\ref{eqn:SEFD_I_error} based on simulated data is shown in Fig.~\ref{fig_mean_ratio_simul}. The trends of high delta in the diagonal planes due to the approximation using equation~\ref{eqn:SEFD_I_phi0_in_XXYY} is similar to that of Fig.~\ref{fig:Hz_dip_delta}.
The $\text{SEFD}_I$ calculated using equation~\ref{eqn:SEFD_I_res} and the approximation using equation~\ref{eqn:SEFD_I_phi0_in_XXYY} agree to within 2.2\% at zenith. However, the approximation is only valid in the cardinal planes. In the diagonal planes, the error reaches 40\% between $\mathrm{ZA=60}\si{\degree}$ and the horizon. The asymmetry in Fig.~\ref{fig_mean_ratio_simul} is due to the mutual coupling from MWA bow-tie antennas surrounding element 6. Next, we describe the process of obtaining the SEFD using observational data, which is an independent method of validation. 

\subsection{Verification with the MWA data}
\label{sec:verification_on_the_mwa_data}

The relative difference defined by equation~\ref{eqn:SEFD_I_error} was verified using 108\,s of MWA data collected in the all-sky observing mode. The data were collected on 2015 February 14 between 14:10:16 and 14:12:04 UTC in a configuration with only dipole 6 
of the sixteen dipoles in every MWA tile enabled and the remaining dipoles terminated. 
The resulting primary beam is sensitive to nearly the entire hemisphere, which was modeled using the 2016 MWA beam model \citep{sokolowski_colegate_sutinjo_2017} with only dipole 6 enabled. In order to verify the predicted $\Delta_{\text{dipole}}$ (equation ~\ref{eqn:SEFD_I_error}) for the short dipole system (Fig.~\ref{fig:Hz_dip_delta}), and more specifically for the simulation of the MWA dipole 6 (Fig.~\ref{fig_mean_ratio_simul}), the $\text{SEFD}_I$ over the entire hemisphere was derived from all-sky images.

\subsubsection{All-sky images}
\label{ref:allsky_imaging}

The MWA data were converted into CASA measurement sets \citep{casa} and downloaded using the MWA All-Sky Virtual Observatory interface \citep{asvo}. In order to perform phase and flux density calibration, an all-sky model was generated using the Positional Update and Matching Algorithm\footnote{\url{https://github.com/JLBLine/srclists}} \citep[PUMA;][]{puma2018} and GaLactic and Extragalactic All-sky MWA catalog \citep[GLEAM;][]{gleam_nhw}.
The \textsc{calibrate} software \citep{offringa-2016}, upgraded with the newest 2016 MWA beam model \citep{sokolowski_colegate_sutinjo_2017} was used to calibrate the all-sky visibilities and generate calibration solutions.
The phases of the resulting calibration solutions were fitted with a linear function, while the amplitudes were fitted with a $5^{th}$ order polynomial; both fits were then applied to the un-calibrated visibilities.
The all-sky images, with 4\,s time resolution, were formed from all correlations products ($XX, YY, XY$ and $YX$) using the \textsc{WSCLEAN}\footnote{\url{https://sourceforge.net/p/wsclean/wiki/Home/}} program \citep{OffMcK14}. The Briggs robust weighting parameter \citep{Briggs_phd_thesis} was set to $-1$, which is optimal for the MWA Phase 1 data with maximum baselines $\approx$3\,km \citep{2013PASA...30....7T} because weighting schemes closer to natural increase the classical confusion noise. The dirty maps were CLEANed with 100000 iterations and a 0.1 Jy threshold. 

The resulting $XX$ and $YY$ images were divided by corresponding images of the beam in X and Y polarizations generated with the 2016 beam model \citep{sokolowski_colegate_sutinjo_2017}, with only dipole 6 enabled.
All the resulting images ($XX, YY, XY$ and $YX$) were converted to Stokes images ($I, Q, U$ and $V$) also using the 2016 beam model. The standard deviation of the noise in the central circular region of 30-pixel radius in Stokes $I$ images was $\approx$0.46\,Jy/beam. The final products resulting from the above procedure were three sets ($XX, YY$ and Stokes $I$) of 28 primary beam corrected all-sky images corresponding to the  $n_t=28$ timesteps (108\,s in total) of the analyzed MWA data.

\subsubsection{Measuring the SEFD from noise in the all-sky images}
\label{ref:allsky_noisemaps}

The $\text{SEFD}_I$ in all directions in the sky was calculated from the standard deviation of difference all-sky images (later referred to as $N$) as $\text{SEFD}_I=\alpha N$, where $\alpha = \sqrt{0.5 \Delta \nu \Delta t}$, $\Delta t=4$\,s is the integration time, and $\Delta \nu = 30.72$\,MHz is the observing bandwidth. The factor 0.5 accounts for the fact that the standard deviation calculated from the  difference images is higher than in the original images by $\sqrt{2}$. The standard deviation of the noise can be calculated in a small region around each pixel in an all-sky image to form an all-sky noise image (it will also be referred to as the noise map). For this procedure to yield the noise originating from $T_{sys}$ only (i.e. due to instrumental and sky noise), we require the contribution from the flux density variations across the region due to the astronomical sources (both point sources and diffuse emission) to be removed. 

Following this approach, the three series (corresponding to $XX$, $YY$ and Stokes $I$) of differences images (between the subsequent $i$-th and $i-1$ image) were generated, resulting in $n_d=27$ difference images in each of $XX$, $YY$ and Stokes $I$ polarizations. The noise calculated in circular regions around pixels in the resulting difference images is distributed around a mean of zero and is not ``contamined'' by the variations of the flux density within the circular regions due to astronomical sources contained inside these regions. The difference images were also visually inspected and verified to have mean value approximately zero and noise-like structure, without any subtraction artefacts, across the entire images. The $XX$ and $YY$ difference images had the characteristic oval-like shape of the primary beam of the MWA dipole in the corresponding polarization in-printed in the standard deviation, which could be best seen in the noise map images with the standard deviation lowest at the center of the images and highest near the horizon. Therefore, the resulting standard deviation is purely due to instrumental and sky noise ($T_{sys}$). 

In an alternative approach, standard deviation of the noise could be calculated from the series of flux density values along the time axis. This way has also been tested by calculating standard deviation (from the interquartile range) for each pixel in the difference images using $n_d$ values along the time axis. The resulting images of standard deviation (in $XX$, $YY$ and Stokes $I$ polarizations) were analyzed in the same way as the adopted approach using the circular regions and led to the same results.

\begin{figure}[t]
\centering
\noindent
  \includegraphics[width=3.25in]{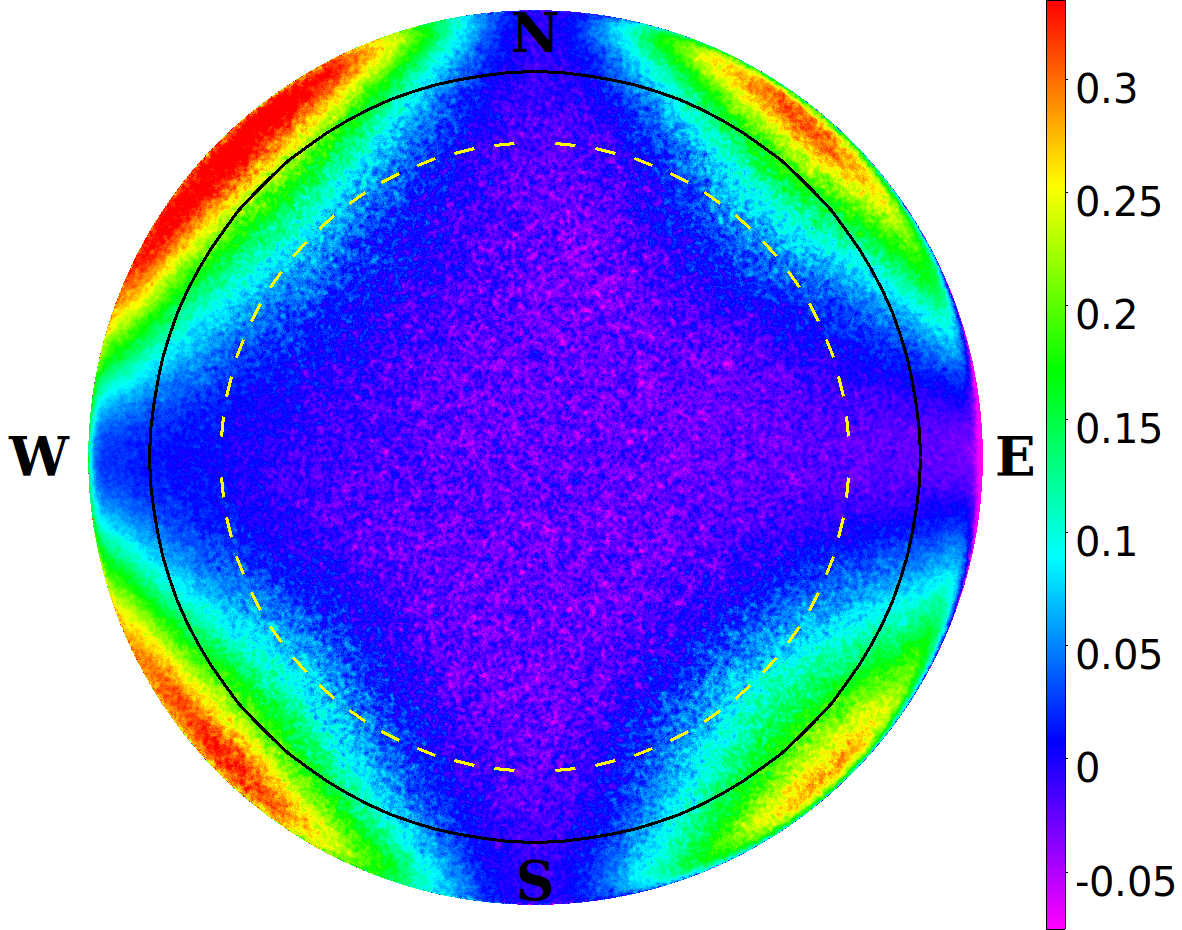}
\caption{The relative difference $\Delta_{\text{dipole}}$ calculated as the median of individual $\Delta_{\text{dipole}}^{i}$ images obtained from the MWA all-sky noise maps using equations~\ref{eqn:ratio} and~\ref{eqn:SEFD_I_error}. The black solid line is at ZA=60 degrees, the yellow dashed line is at ZA=45 degrees, and the letters N, E, S, W show cardinal directions.}
\label{fig_mean_ratio_data}
\end{figure}

The 2D noise maps over the entire hemisphere were calculated as follows. For each of the resulting difference images, the noise at each position on the sky (i.e. around each image pixel) was calculated as the standard deviation of all pixels within a radius of 30 pixels ($\approx$ 91 synthesized beams) from the pixel being analyzed. The interquartile range divided by 1.35 was used in order to be more robust against any outlier data points due to radio-frequency interference (RFI), or residuals of astronomical sources in difference images. 
This procedure was applied to $n_d$ all-sky difference images resulting in $n_d$ noise maps ($N_{I}$, $N_{XX}$ and $N_{YY}$) for each of the polarizations (Stokes $I$, $XX$ and $YY$ respectively).
These noise maps were used to calculate $\Delta_{\text{dipole}}$ defined in equation~\ref{eqn:SEFD_I_error}, by substituting $\text{SEFD}_I=\alpha N$ and with the constant $\alpha$ canceling out, resulting in the following equation:
\begin{eqnarray}
\Delta_{\text{dipole}}^{i} = \frac{( N_{I} - \frac{1}{2}\sqrt{N_{XX}^2 + N_{YY}^2} )}{N_{I}},
\label{eqn:ratio}
\end{eqnarray}
where $i$ is the image index. For each of the difference images ($n_d$ in total), the corresponding noise maps ($N_{I}$, $N_{XX}$ and $N_{YY}$) were used to calculate $n_d$ number of $\Delta_{\text{dipole}}^{i}$ images according to equation~\ref{eqn:ratio}. Visual inspection of these images revealed that although they were very similar to Figures~\ref{fig:Hz_dip_delta} and \ref{fig_mean_ratio_simul}, the centers were quite noisy due to near zero values. Therefore, in order to reduce the noise, a median image $\Delta_{\text{dipole}}^{\text{data}}$ was calculated out of $n_d$ $\Delta_{\text{dipole}}^{i}$ images and is shown in Figure~\ref{fig_mean_ratio_data}. 

\subsection{Comparison between data and simulations}
\label{sec:compare_data_vs_simul}

\begin{figure}[t]
\centering
\noindent
  \includegraphics[width=3.4in]{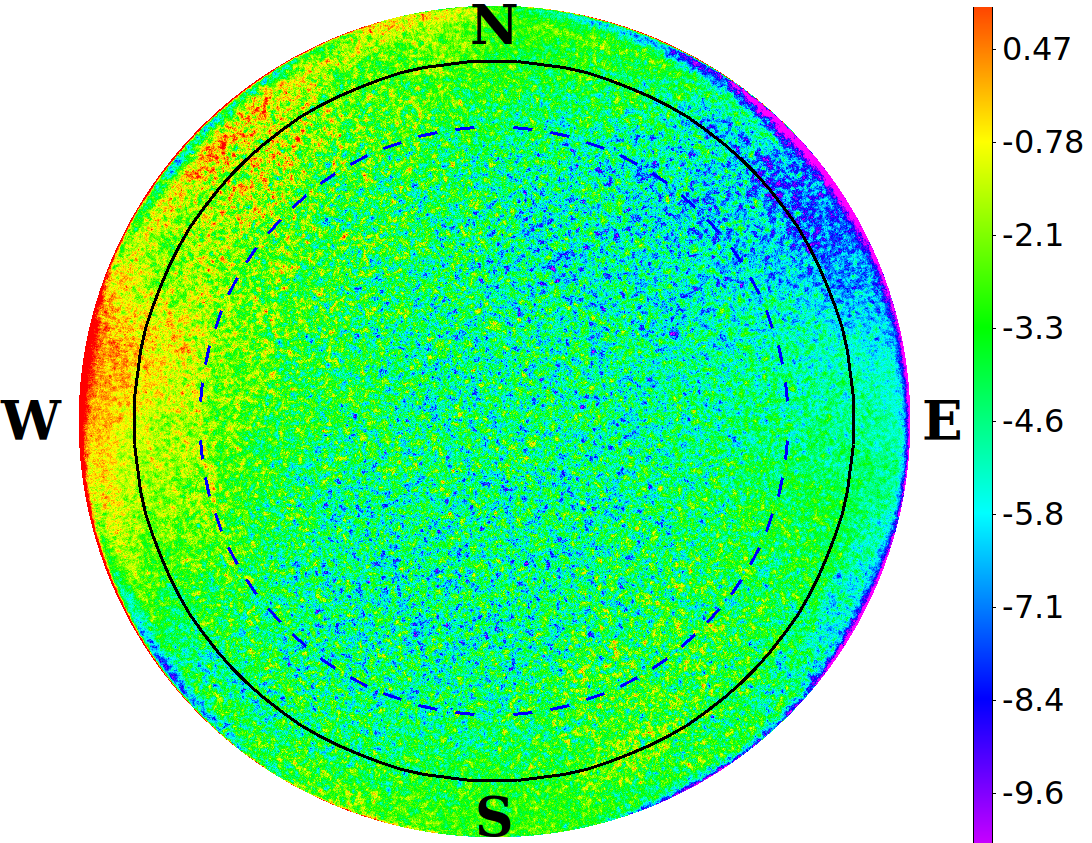}
\caption{The percentage difference of the MWA data $\Delta_{\text{dipole}}^{\text{data}}$ (Fig.~\ref{fig_mean_ratio_data}) and the prediction from the simulation $\Delta_{\text{dipole}}^{\text{sim}}$ (Fig.~\ref{fig_mean_ratio_simul}). The black solid line is at ZA=60 degrees, the blue dashed line is at ZA=45 degrees, and the letters N, E, S, W show cardinal directions.}
\label{fig_diff_percent}
\end{figure}

Figure~\ref{fig_mean_ratio_data} is very similar to Figure~\ref{fig:Hz_dip_delta}, and nearly identical to  Figure~\ref{fig_mean_ratio_simul} showing the $\Delta_{\text{dipole}}^{\text{sim}}$ expected from the MWA simulations of the same observing setup. $\Delta_{\text{dipole}}^{\text{sim}}$ is defined as $\Delta_{\text{dipole}}$ (equation~\ref{eqn:SEFD_I_error}) predicted by the MWA simulation (Sec.~\ref{sec:mwa_simulation}) in order to unequivocally distinguish it from the corresponding difference obtained from the MWA data, which is denoted as $\Delta_{\text{dipole}}^{\text{data}}$. Similarly, $\text{SEFD}_I^{\text{sim}}$ and $\text{SEFD}_I^{\text{data}}$ are defined as the SEFD calculated from the MWA simulation and data respectively. Figure~\ref{fig_diff_percent} shows the percentage difference between the $\Delta_{\text{dipole}}^{\text{data}}$ derived from the MWA data (Fig.~\ref{fig_mean_ratio_data}) and $\Delta_{\text{dipole}}^{\text{sim}}$ predicted by the simulations (Fig.~\ref{fig_mean_ratio_simul}). The absolute values of the differences are within a few percent in the center (and to elevations as low as approximately 30\degree), reaching the largest values (exceeding $|10$\%$|$) only near the horizon, where the MWA beam model becomes less reliable.

\begin{figure}[t]
\centering
\noindent
  \includegraphics[width=3.25in]{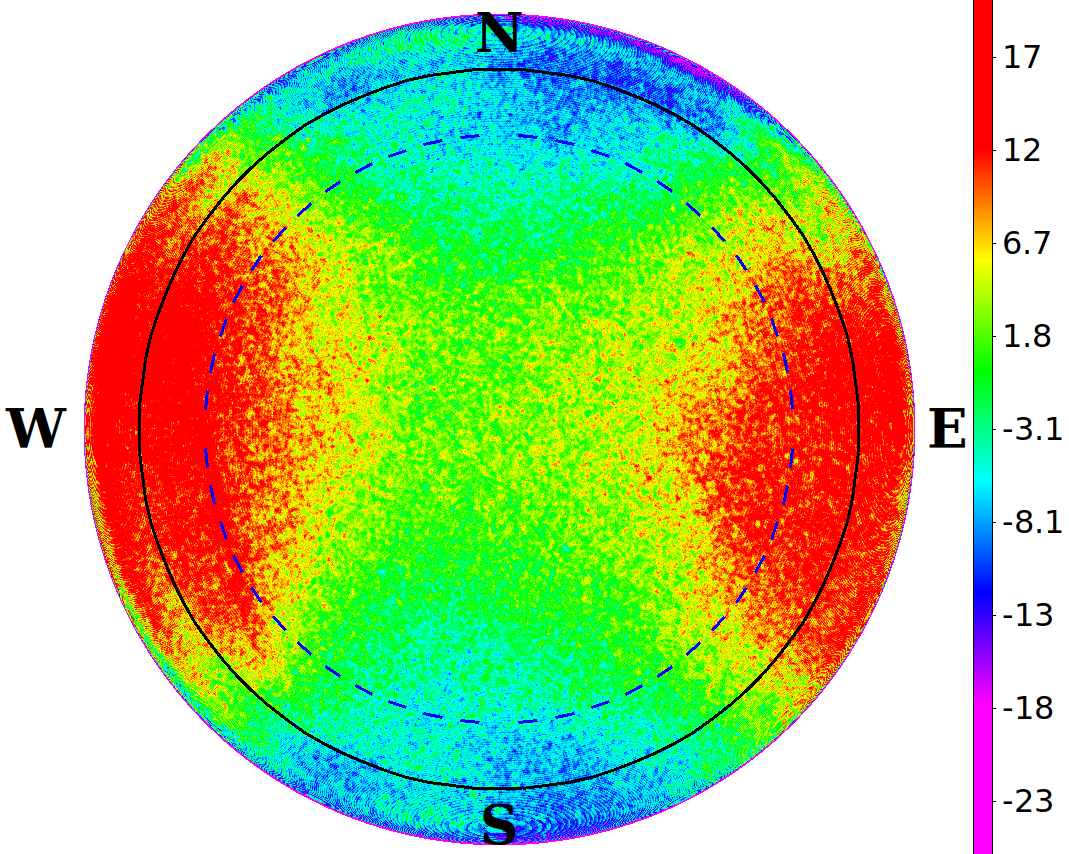}
\caption{The relative difference between the data and simulations, $\Delta_{\text{SEFD}}$ (in percent), calculated as $(\text{SEFD}_I^{\text{data}}- \text{SEFD}_I^{\text{sim}})/\text{SEFD}_I^{\text{sim}} \times 100$\%. The agreement between the data and simulation at zenith is remarkable with measured sensitivity $\text{SEFD}_I^{\text{data}}$(ZA=0\degree)\,$= 463700 \pm 200$\,Jy and simulated $\text{SEFD}_I^{\text{sim}}$(ZA=0\degree)\,$= 463250 \pm 20$\,Jy ($\approx$0.1\% difference). Both values ($\text{SEFD}_I^{\text{data}}$ and $\text{SEFD}_I^{\text{sim}}$) were calculated as the median and interquartile range based standard deviation in a 10-pixel radius around the center. The black solid line is at ZA=60 degrees, the blue dashed line is at ZA=45 degrees, and the letters N, E, S, W show cardinal directions.}
\label{fig_diff_sefd_percent}
\end{figure}

In a similar way, the median $\text{SEFD}_I^{\text{data}}$ image was obtained from the data as the median of $n_d$ individual $\text{SEFD}_I^{i}$ images calculated from the noise maps as $\text{SEFD}_I^{i}=\alpha N$, and it was compared with the $\text{SEFD}_I^{\text{sim}}$ derived from the MWA simulations. Figure~\ref{fig_diff_sefd_percent} shows the relative difference between the data and simulations, calculated as $\Delta_{\text{SEFD}} = (\text{SEFD}_I^{\text{data}} - \text{SEFD}_I^{\text{sim}})/\text{SEFD}_I^{\text{sim}} \times 100$\%. The agreement between the data and simulations at zenith is remarkable with measured sensitivity $\text{SEFD}_I^{\text{data}}$(ZA=0\degree)\,$= 463700 \pm 200$\,Jy and simulated $\text{SEFD}_I^{\text{sim}}$(ZA=0\degree)\,$= 463250 \pm 20$\,Jy, which means the difference is $\approx$0.1\%. Both values ($\text{SEFD}_I^{\text{data}}$ and $\text{SEFD}_I^{\text{sim}}$) were calculated as the median and interquartile range based standard deviation in 10 pixels radius around the center. 
These differences between the data and simulations remain mostly within $\pm5$\% above an elevation $45$\degree and increase to $\pm20$\% only below an elevation of 30\degree. The increasing discrepancy between simulations and observational measurements at lower elevations reflects the fact that the FEKO model is an approximation, which generally becomes less accurate the further the beam is from zenith. This is consistent with prior findings in \cite{sokolowski_colegate_sutinjo_2017} regarding the beam model accuracy.

These images provide the evidence that the data and simulations are in a very good agreement and demonstrate correctness of the findings presented in this paper.

\section{Conclusion}
\label{sec:concl}
There are a few important conclusions. First, the SEFD is the proper figure of merit for a polarimetric interferometer, but \emph{not} as commonly defined using $A_e/T_{sys}$ one antenna polarization at a time. Rather, the SEFD for the dual-polarized antenna system is directly computable through the standard deviation of the flux density estimate over the entire sky with no assumption regarding the polarization state of the source. This is given by the formula in equation \ref{eqn:SEFD_I_res}.  Second, the SEFD defined in this way produces one number for the dual-polarized antenna system such that the sensitivity at off-zenith angles can be calculated with no ambiguity (see Sec.~\ref{sec:SKA1Low}). Finally, the SEFD formula calculated based on full-wave simulation was validated with the MWA to a very good agreement using all-sky astronomical observations and electromagnetic simulations. The difference between the two are within the noise level of the observation in one case, and in another, within known accuracy levels of electromagnetic simulation for far off-zenith angles.  

\section*{Acknowledgement}
\label{sec:Ack}
The authors thank A/Prof C. Trott for feedback and discussions on the statistical calculations.

\section*{Appendix: Treatment of statistics}
The noise voltages in Fig.~\ref{fig:antenna_noise} are thermal noise sources which are zero-mean and Gaussian distributed. Our treatment uses complex envelope quantities; the real and imaginary components of the thermal noise voltages are independent and identically distributed (iid). The reasoning in Sec.~\ref{sec:STdev_Calc} leads to Fig.~\ref{fig:indep_noise} in which every noise source is independent. Therefore in our treatment, it is adequate to address only the statistics of independent and zero-mean Gaussian noise sources. In equation~\ref{eqn:eeh11_22} and equation~\ref{eqn:var_11}, we see a sum of random variables as follows
\begin{eqnarray}
W = aX_1X_2^*-zX_1Y_2^* -z^*Y_1X_2^*+bY_1Y_2^*,
\label{eqn:sum}
\end{eqnarray}
where $X_1,X_2, Y_1, Y_2$ are complex random variables; $a, b$ are real constants and $z$ is a complex constant.
\begin{eqnarray}
\text{Var}(W) &=& \text{Var}(aX_1X_2^*-zX_1Y_2^* -z^*Y_1X_2^*+bY_1Y_2^*),
\nonumber \\
&=&|a|^2\text{Var}(X_1X_2)+|z|^2\text{Var}(X_1Y_2)+\nonumber\\
&+&|z|^2\text{Var}(Y_1X_2)+|b|^2\text{Var}(Y_1Y_2)+2C,
\label{eqn:varW}
\end{eqnarray}
where $C$ is the covariance of cross terms.
\begin{eqnarray}
C &=& -az\text{Cov}(X_1X_2^*,X_1Y_2^*)-az^*\text{Cov}(X_1X_2^*,Y_1X_2^*) 
\nonumber\\
&+&ab \text{Cov}(X_1X_2^*,Y_1Y_2^*) +|z|^2\text{Cov}(X_1Y_2^*,Y_1X_2^*)
\nonumber \\
&-&bz\text{Cov}(X_1Y_2^*,Y_1Y_2^*)-bz^*\text{Cov}(Y_1X_2^*,Y_1Y_2^*).
\label{eqn:C}
\end{eqnarray}
Because of independent noise, $\text{Cov}(X_1X_2^*,Y_1Y_2^*)=\text{Cov}(X_1Y_2^*,Y_1X_2^*)=0$, which leaves $\text{Cov}(X_1X_2^*,X_1Y_2^*)$ and similar terms in which there is a common $X_1$ term separated by the comma $(X_1\_,X_1\_)$, similarly $\text{Cov}(X_1X_2^*,Y_1X_2^*)$, $\text{Cov}(X_1Y_2^*,Y_1Y_2^*)$, $\text{Cov}(Y_1X_2^*,Y_1Y_2^*)$.
\begin{eqnarray}
\text{Cov}(X_1X_2^*,X_1Y_2^*)&=&\left<X_1X_2^*X_1Y_2^*\right>-\left<X_1X_2^*\right>\left<X_1Y_2^*\right> \nonumber \\
&=&\left<X_1X_2^*X_1Y_2^*\right>,
\label{eqn:X1X2,X1Y2}
\end{eqnarray}
where the last line is due to independent zero-mean noise sources such that $\left<X_1X_2^*\right>=\left<X_1Y_2^*\right>=0$. To find $\left<X_1X_2^*X_1Y_2^*\right>$, we use a key formula for $Z_{1,2,3,4}$ that are zero-mean joint Gaussian random variables~\citep{Thompson2017_rx_sys, Baudin_App3}
\begin{eqnarray}
\left<Z_1Z_2Z_3Z_4\right>&=&\left<Z_1Z_2\right>\left<Z_3Z_4\right>
+\left<Z_1Z_3\right>\left<Z_2Z_4\right>\nonumber \\
&+&\left<Z_1Z_4\right>\left<Z_2Z_3\right>,
\label{eqn:Thompson_formula}
\end{eqnarray}
which is valid for real and complex zero-mean joint Gaussian random variables. Applying this to equation~\ref{eqn:X1X2,X1Y2}, we get
\begin{eqnarray}
\left<X_1X_2^*X_1Y_2^*\right>&=&\left<X_1X_2^*\right>\left<X_1Y_2^*\right>
+\left<X_1X_1\right>\left<X_2^*Y_2^*\right>\nonumber \\
&+&\left<X_1Y_2^*\right>\left<X_2^*X_1\right>\nonumber \\
&=&0,
\label{eqn:Thompson_formula_applied}
\end{eqnarray}
which is zero because of zero-mean independent noise. This is similarly the case for $\text{Cov}(X_1X_2^*,Y_1X_2^*)$, $\text{Cov}(X_1Y_2^*,Y_1Y_2^*)$, $\text{Cov}(Y_1X_2^*,Y_1Y_2^*)$. In conclusion, $2C=0$ in equation~\ref{eqn:varW}.


\end{document}